\documentclass[journal]{IEEEtran}
\ifCLASSINFOpdf

\else

\fi
\usepackage{graphicx}
\usepackage{subfig}

\usepackage{bm}
\usepackage{amssymb}
\usepackage{amsmath}
\usepackage{footnote}
\usepackage{footmisc}
 
\usepackage{amsfonts}
\usepackage{textcomp}
\usepackage{xcolor}
\usepackage{cite}
\usepackage{booktabs}
\usepackage{url}
\usepackage{hyperref}
\usepackage{color}  
\usepackage{multirow}

\begin{document}
\title{Learned Block-based Hybrid Image Compression}

\author{Yaojun Wu$^*$, Xin Li$^*$, Zhizheng Zhang, Xin Jin and Zhibo Chen$^\dagger$,~\IEEEmembership{Senior Member,~IEEE}
\thanks{This work was supported in part by NSFC under Grant U1908209, 61632001, 62021001 and the National Key Research and Development Program of China 2018AAA0101400}
\thanks{Yaojun Wu, Xin Li, Zhizheng Zhang, Xin Jin, and Zhibo Chen are with the Department of Electronic Engineer and Information Science, University of Science and Technology of China, Hefei, Anhui, 230026, China (e-mail: yaojunwu@mail.ustc.edu.cn; lixin666@mail.ustc.edu.cn;  zhizheng @mail.ustc.edu.cn; jinxustc@mail.ustc.edu.cn; chenzhibo@ustc.edu.cn).} 
}
\markboth{IEEE Transactions on Circuits and Systems for Video Technology}%
{Shell \MakeLowercase{\textit{et al.}}: Bare Demo of IEEEtran.cls for IEEE Journals}

\maketitle
\footnotetext[1]{Yaojun Wu and Xin Li contribute equally to this paper.} 
\footnotetext[2]{Zhibo Chen is the corresponding author.} 
\begin{abstract}
Recent works on learned image compression perform encoding and decoding processes in a full-resolution manner, resulting in two problems when deployed for practical applications. 
First, parallel acceleration of the autoregressive entropy model cannot be achieved due to serial decoding. Second, full-resolution inference often causes the out-of-memory (OOM) problem with limited GPU resources, especially for high-resolution images. Block partition is a good choice to handle the above issues, but it brings about new challenges in reducing the redundancy between blocks and eliminating block effects. To tackle the above challenges, this paper provides a learned block-based hybrid image compression (LBHIC) framework. Specifically, we introduce explicit intra prediction into a learned image compression framework to utilize the relation among adjacent blocks. Superior to context modeling by linear weighting of neighbor pixels in traditional codecs, we propose a contextual prediction module (CPM) to better capture long-range correlations by utilizing the strip pooling to extract the most relevant information in neighboring latent space, thus achieving effective information prediction. Moreover, to alleviate blocking artifacts, we further propose a boundary-aware postprocessing module (BPM) with the edge importance taken into account. Extensive experiments demonstrate that the proposed LBHIC codec outperforms the VVC, with a bit-rate conservation of 4.1\%, and reduces the decoding time by approximately 86.7\% compared with that of state-of-the-art learned image compression methods.
\end{abstract}

\begin{IEEEkeywords}
Learned image compression, block-based, prediction, postprocessing, acceleration.
\end{IEEEkeywords}

%
\IEEEpeerreviewmaketitle

\section{Introduction}

\IEEEPARstart{I}{mage} compression is an essential technique to reduce the costs of image transmission and storage. Traditional image codecs, such as JPEG~\cite{wallace1992jpeg}, BPG~\cite{bellard2015bpg}, and VVC (intra)~\cite{bross2018versatile}, adopt a hybrid coding framework consisting of prediction, transformation, quantization, and entropy coding. However, traditional image codecs are limited by handcrafted prediction modes and lack adaptability. Owing to significant progress in artificial neural networks, some works have attempted to utilize  CNN to replace the part of traditional codecs \cite{schiopu2019cnn}, which is still limited to the handcrafted architecture. Meanwhile, learned image compression methods have also been proposed based on transformation coding with an automatic end-to-end optimization \cite{balle2017end}.

Learned image compression methods~\cite{balle2017end,balle2018variational, minnen2018joint,johnston2018improved,johnston2019computationally,sun2020semantic,chen2019learning, zhang2019learned, guo2021causal, mentzer2020high}  perform encoding and decoding in a full-resolution manner, which commonly consists of transformation, quantization, and entropy modeling. Among these modules, probabilistic-based entropy modeling is an essential component to estimate rate for rate-distortion optimization. Current advanced entropy modeling achieves this aim through hyperprior modeling~\cite{balle2018variational} and autoregressive context modeling \cite{minnen2018joint}. The autoregressive context module estimates the probability based on the previously decoded contents in the latent space, which can achieve a bit-rate conservation of 15.87\%\cite{minnen2018joint}. The context modeling confers important benefits in compression but also greatly increases the time complexity, especially in the decoding process. The probability for the content to be decoded is estimated according to the already-decoded contents. Thus, the entire decoding step must be processed in sequence, which is severely time-consuming. In addition, the required memory of the GPU resources for the frame-level coding unit will dynamically increase as the resolution of the input image increases, which often causes the out-of-memory (OOM) problem with limited GPU resources.

\begin{figure*}[t]
    \centering
    \includegraphics[width=0.95\linewidth]{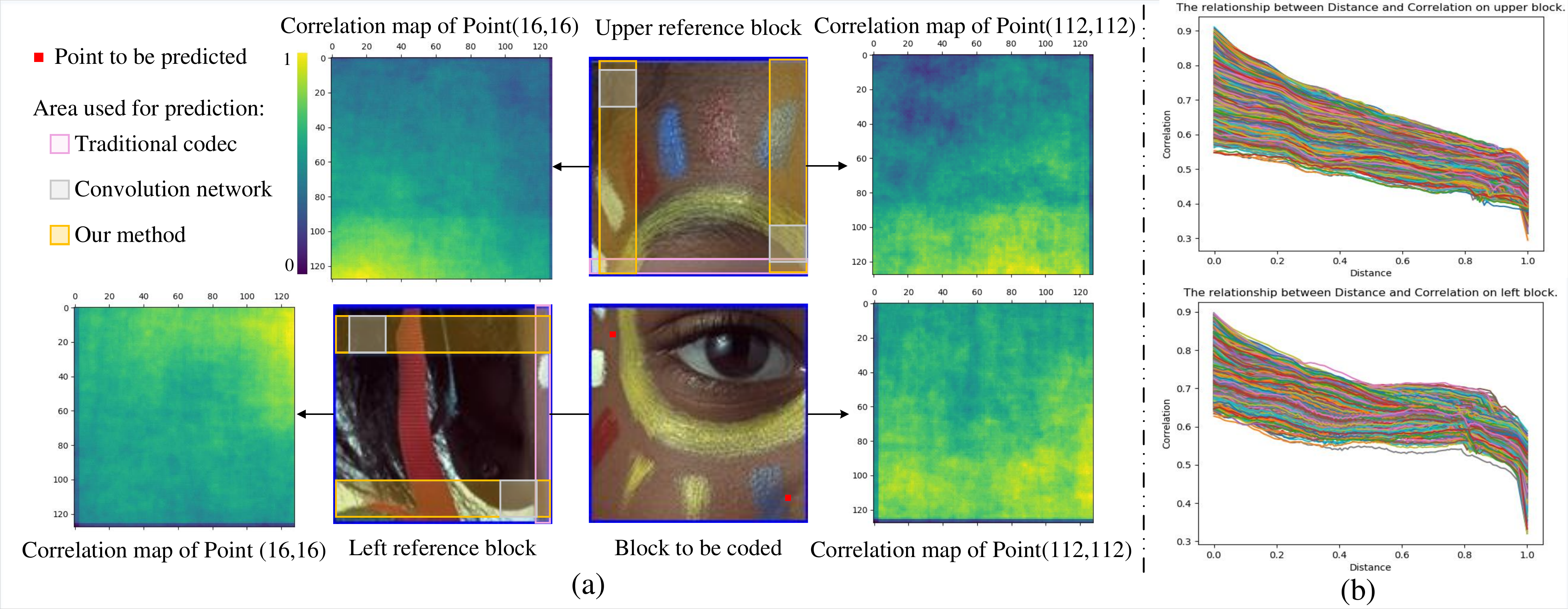}
    \caption{Example to illustrate why our CPM can extract more relevant information than the prediction modules of the traditional codecs and convolution network. 
    (a) The visualization of corresponding areas for different prediction methods. (b)
   The statistical relationship between the distance and the correlation by calculating the correlation and distance of all pixels to be predicted and all pixels in reference blocks.}
    \label{fig:first}
\end{figure*}

Considering the coding speed and limited GPU resources, we are in pursuit of block-based coding frameworks that allow us to perform encoding-decoding in a parallel manner. However, based on our experimental observations, block partition commonly incurs compression performance degradation due to the following reasons:
First, the relevant information among blocks cannot be utilized efficiently since adjacent blocks are coded independently.
Second, the padding operation in the boundaries of the convolution affects the reconstruction of edge pixels, resulting in block artifacts. In addition, explicit intra prediction is easier to be designed for block-based coding frameworks relative to full-resolution compression frameworks.

This paper aims to solve the above limitations and propose an effective and efficient learned image compression framework.
Unlike existing learned compression methods based on transformation coding, we propose a learned block-based hybrid image compression (LBHIC) method, which introduces a contextual prediction module (CPM) to utilize the relationship between adjacent blocks, and propose a boundary-aware postprocessing module (BPM) to remove the block artifacts. 

For better prediction among adjacent blocks, we analyze the correlation map between all pixels of the predicted block and reference blocks. As shown in Fig.~\ref{fig:first}(a), to compare the advantages of different prediction methods, we randomly give two points as an example to demonstrate the reference areas of different prediction methods. Meanwhile, we also calculate the correlation map of predicted points and all pixels in reference blocks. According to Fig.~\ref{fig:first}(a), our CPM can captures the most relevance information from the reference block to prediction block compared with other methods. To provide the statistical analysis, we evaluate  the relation between the distance and the correlation of all predicted pixels and reference pixels from whole kodak dataset \cite{franzen1999kodak}, 
which can be seen in Fig.~\ref{fig:first}(b). In these two plots of Fig.~\ref{fig:first}(b), each color line represents one pixel to be predicted. We calculate the distance between the predicted pixel and all pixels in reference blocks by Euclidean distance, and utilize mean-normalization to normalize the distance in order to show the relative distance between different points to be predicted and all points of reference blocks. Based on the statistical results of correlation distribution in the correlation map, we find that the positions of the pixel to be predicted and the most relevant pixel in the reference block are not matched and that their correlation is inversely related to their distance. Nevertheless, traditional codecs utilize the linear weighting of neighboring pixels to obtain the prediction of the current pixel (e.g., the pink mask in Fig.~\ref{fig:first}(a)), which only utilizes limited pixels and thus is difficult to optimize in an end-to-end manner. Another straightforward solution is to feed reference blocks directly into a convolution network for prediction. However, the correlation between the target point and its relevant context depends on their relative positions. In addition, this correlation may be long range. Thus, the convolution-based method encounters difficulty in extracting the most relevant information for prediction (e.g., the gray mask in Fig.~\ref{fig:first}(a)).

Different from the above schemes, our CPM captures most of the relevant information through the operation of strip pooling~\cite{hou2020strip}, which is shown as the yellow mask in Fig.~\ref{fig:first}(a). 
Since the upper and left blocks have different spatial relationships with the points to be predicted, the distribution of the most relevant pixels is different. For the upper block, the most relevant information is located in the bottom part, while the right part is the most relevant part for the left block.
Therefore, we extract relevant information through the vertical/horizontal band for prediction, respectively. Next, we fuse the information obtained from the upper and left reference blocks and feed it to a network to realize the final prediction.

To further improve the subjective and objective quality of the coded image, we also propose a boundary-aware postprocessing module (BPM) to cope with the blocking artifacts caused by the influence of padding in the network. 
Specifically, we first locate the boundary between different coded blocks of the image and then generate a boundary mask by expanding the boundary. Afterwards, we utilize the boundary mask to guide the postprocessing network to focus more on the boundary area and remove the coding artifacts, especially the edges across different blocks. Moreover, the coding artifacts also contain multiple distortions (blur, noise, etc.). Here, to improve the representation ability of the network, our BPM consists of a multiscale architecture and revised grouped dense residual blocks. Owing to our proposed BPM, we can remove the block effect and achieve approximately 0.3 dB gain over our basic block-based image compression network.

The contributions of this paper are summarized as follows:
\begin{itemize}
\item We design learned block-based hybrid image compression (LBHIC). This approach takes advantage of the traditional hybrid coding method and learned image compression and compensates for their respective drawbacks.

\item We analyze the correlation of the adjacent blocks and propose a contextual prediction module (CPM) to achieve long-range relevance capture, which yields more effective information decorrelation, thus achieving better rate-distortion performance.

\item We further propose the boundary-aware postprocessing module (BPM) to remove artifacts and achieve better subjective and objective quality.

\item Extensive experiments validate that the proposed LBHIC achieves state-of-the-art (SOTA) compression performance compared with other image compression methods in terms of PSNR, MS-SSIM, and coding speed.
\end{itemize}

\begin{figure*}[t]
    \centering
    \includegraphics[width=0.80\linewidth]{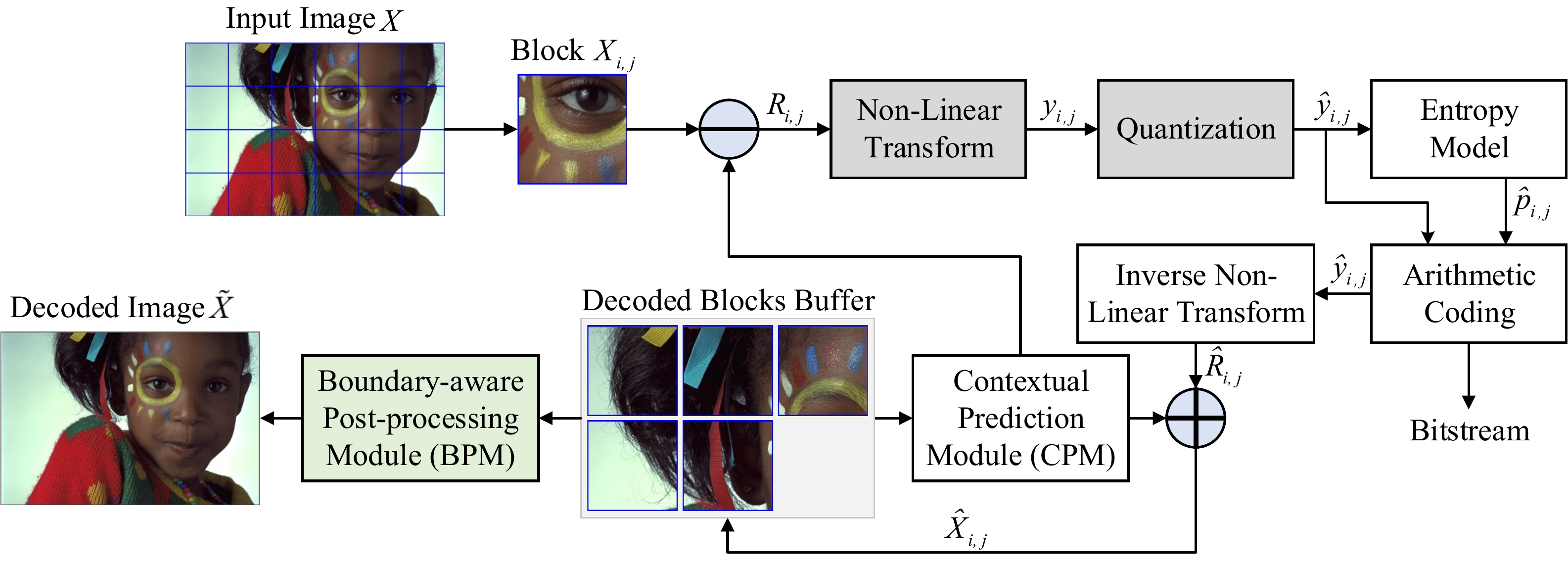}
    \caption{Flowchart of the proposed LBHIC. Modules in gray are only included on the encoder side, while the module in green is only contained on the decoder side.}
    \label{fig:framework}
\end{figure*}

The remainder of this paper is organized as follows. Section~\ref{sec:2} introduces work related to our endeavor. In Section~\ref{sec:3}, we formulate our LBHIC framework and then provide a detailed description of our model. We will present experiment results in Section~\ref{sec:4} and conclude with Section~\ref{sec:5}.

\section{Related Work}
\label{sec:2}
Lossy image compression standard codecs have been developed for decades. As one of the most widely used standards, JPEG was first created in 1992. Subsequently, to improve compression performance, standards such as JPEG2000, WebP, BPG, and VVC (intra) were successively proposed. To our knowledge, VVC (intra) offers the highest compression performance among standard codecs.

Recently, learned image compression has attracted great attention to achieve improved compression performance in this field. From the perspective of neural topology, learned image compression methods can be divided into two categories: recurrent models and variational autoencoder (VAE) models. Recurrent models~\cite{toderici2015variable, toderici2017full, johnston2018improved} compress the image by iteratively feeding the residual information (between the reconstructed image and the original image) into the encoder. In the framework, the codec can realize variable rates through a single network without any retraining process. 
VAE models \cite{balle2017end, balle2018variational, johnston2019computationally} are composed of nonlinear transformation, relaxed quantization, entropy coding and postprocessing, which are modified in following works to further improve compression performance.

\noindent\textbf{Nonlinear Transformation:}
To better remove the spatial redundancy of images, GDN~\cite{balle2016end} and nonlocal attention~\cite{chen2019neural} are proposed to perform nonlinear transformation, removing local and global redundancy, respectively. In addition, Qian et al.~\cite{qian2020learning} go one step further: they introduce the subtraction operation in GDN to correct the mean-shifting problem.

\noindent\textbf{Relaxed Quantization:}
Unlike traditional codecs, a rounding operation is not differentiable and thus is not suitable for the quantization in learning-based compression. To develop differentiable quantization, Ball\'{e}~\cite{balle2017end} utilizes the additive uniform noise as the substitution of the round operation during the training, while Agustsson et al.~\cite{agustsson2017soft} propose a substitution of the vector quantization for better performance. Furthermore, to eliminate the mismatch between training and testing phases in quantization, universal quantization~\cite{agustsson2020universally} is introduced in recent works.

\noindent\textbf{Entropy coding:}
Entropy coding is an essential part of the VAE-based framework and serves the role of estimating rates during rate-distortion optimization, which compresses the discrete latent representation into a bit-stream in a lossless manner. To improve the performance of entropy coding, a probabilistic model~\cite{balle2018variational} is first proposed, which uses a hyperprior to capture the dependencies in the latent space. Minnen et al., Mentzer et al., and Lee et al.~\cite{minnen2018joint, mentzer2018conditional, Lee2019Context} utilize the autoregressive model to establish the context relationship between coded elements and the element to be coded in the spatial latent space in order to further improve the accuracy of probability estimation. 
In addition, a 3D-autoregressive model~\cite{guo20203} is considered for the same purpose, which combines the spatial and channel relationships to further capture the relationship within the elements. To further improve the long-term dependency in the probabilistic model, Hu et al.~\cite{hu2020coarse} propose hierarchical hyperprior layers to effectively reduce spatial redundancy. 

\noindent\textbf{Postprocessing:}
With the development of image restoration technologies, such as image denoising \cite{dabov2007image, zhang2017beyond}, image deblurring \cite{kupyn2018deblurgan, kupyn2019deblurgan} and hybrid-distorted image restoration technologies \cite{li2020learning}, postprocessing has been applied in image compression~\cite{lee2019hybrid, jin2020dual}. Lee et al.~\cite{lee2019hybrid} proposed to jointly train the image compression network and postprocessing module to reduce the artifacts caused by compression. To further improve the subjective quality of image compression, some studies \cite{kim2020towards, li2020multi} utilize perceptual loss \cite{wang2018esrgan} and adversarial loss \cite{wang2018esrgan} to guide the training process of the postprocessing network. 

Although the above schemes yield better compression performance, coding images with full resolution ignores the compression complexity. To accelerate coding time, Johnston et al.~\cite{johnston2019computationally} utilize automatic network optimization to reduce the computational complexity of the network. Another solution is to utilize block partition for compression. 
The block partition was first applied in the learning based video coding scheme proposed by Chen et al.~\cite{chen2019learningvideo}, which utilized motion extension and hybrid prediction networks to model block-level spatiotemporal coherence, but its intra coding scheme still functions in a full-resolution manner. 
For image compression, Lin et al.~\cite{lin2020spatial} propose a block-based scheme for acceleration and utilize RNN-based transformation to reduce the redundancy between blocks by utilizing a copy-paste-like operation. This approach is similar to the convolution network in Fig.~\ref{fig:first}, which lacks the utilization of the most relevant information and lacks applicability in most scenarios. 

\section{Methodology}
\label{sec:3}
\subsection{Overview of our Framework}
To reconcile compression efficiency with compression performance, we propose a learned block-based hybrid image compression framework (LBHIC), which integrates the predictive coding to remove the redundancy among adjacent blocks and introduces the postprocess module to cope with block artifacts. The overall framework is shown in Fig.~\ref{fig:framework}, containing block partition, nonlinear transform, quantization, entropy model, arithmetic coding, inverse nonlinear transform, contextual prediction module (CPM) and boundary-aware postprocessing module (BPM).

Specifically, we first divide the input image $\bm{X}$ into nonoverlapping blocks as shown in Fig.~\ref{fig:framework}. Then we utilize predictive coding to capture and reduce the redundancy between the current block $\bm{X}_{i, j}$ and adjacent decoded blocks from the decoded block buffer.

Here, we denote the prediction network as $g_{pred}$ and its parameters as $\bm{\theta}_{pred}$. $\bm{X}_{i, j}$ denotes the block in the $i$-th row and $j$-th column, and $\bm{\hat{X}}_{i, j}$ denotes the corresponding decoded block. With the prediction of $\bm{X}_{i, j}$ generated by CPM, we can calculate the residual information $\bm{R}_{i, j}$ as:
\begin{equation}
\bm{R}_{i, j} = \bm{X}_{i, j}-g_{pred}(\bm{\hat{X}}_{i-1, j},\bm{ \hat{X}}_{i, j-1}; \bm{\theta}_{p}).
\label{eq:predict}
\end{equation}

After prediction, we follow the structure of the VAE-based image compression method to compress the residual information. This approach utilizes a nonlinear transformation network $g_{nlt}$ with parameters $\bm{\theta}_{nlt}$ to map the residual information $\bm{R}_{i, j}$ into a latent representation $\bm{y_{i, j}}$, which can reduce the spatial redundancy inside the block. We then utilize quantization module $Q$ to discretize the value of the latent variable $\bm{y_{i, j}}$, which can reduce the information in a lossy way. The above process can be formulated as:
\begin{equation}
\bm{\hat{y}}_{i, j} = Q(\bm{y}_{i, j})=Q(g_{nlt}(\bm{R}_{i, j}; \bm{\theta}_{nlt})).
\label{eq:quantized}
\end{equation}

Similar to Minnen et al.~\cite{minnen2018joint}, we estimate the probability distribution of the quantized elements for more accurate rate estimation in order to improve the compression performance. Let $\hat{y}_{i, j, k}$ denote the $k$-th element in the quantized latent feature, where its probability distribution $\hat{p}_{i, j, k}$ is described by the Gaussian mixture model (GMM):
\begin{equation}
\hat{p}_{i, j, k} = \prod_{t=1}^{N}\frac{\alpha_{i, j, k, t} }{\sqrt{2\pi\sigma_{i, j, k, t}^2}}exp(-\frac{(y_{i, j, k}-\mu_{i, j, k, t})^2}{2\sigma_{i, j, k, t}^2}),
\label{eq:GMM}
\end{equation}
where the parameters $\bm{\theta}_{GMM}$ ($\alpha$, $\mu$, and $\sigma$) of the GMM are estimated by a parametric entropy coding model $g_{e}(\bm{\hat{y}}_{i, j}; \bm{\theta}_{e})$. Specifically, in our LBHIC, the entropy coding model contains a hyperprior model~\cite{balle2018variational} and autoregressive model~\cite{minnen2018joint}. Based on $\hat{p}_{i, j, k}$, we convert the quantized element $\hat{y}_{i, j, k}$ into bit-streams through arithmetic coding.

To obtain the reconstructed block $\bm{\hat{X}}_{i, j}$, we recover the residual information through the inverse nonlinear transformation network $g_{inlt}$ with parameters $\bm{\theta}_{inlt}$. We then reconstruct the block to be compressed according to the recovered residual and the predicted content. As a result, the reconstructed block $\bm{\hat{X}_{i, j}}$ is obtained as follows:
\begin{equation}
\bm{\hat{X}_{i, j}} = g_{inlt}(\bm{\hat{y}}_{i, j}; \bm{\theta}_{inlt})+pred(\bm{\hat{X}}_{i-1, j},\bm{ \hat{X}}_{i, j-1}; \bm{\theta}_{p}).
\label{eq:decode}
\end{equation}

On the decoder side, we first reconstruct all blocks and then assemble them into a full resolution reconstructed image $\bm{\hat{X}}$. To eliminate the block artifacts caused by the block-based compression scheme, we further use postprocessing network $g_{post}$ with parameters $\bm{\theta_{post}}$ to improve subjective and objective quality, which is formulated as:
\begin{equation}
\bm{\tilde{X}} = g_{post}(\bm{\hat{X}}; \bm{\theta}_{post}).
\label{eq:post}
\end{equation}

Details about the prediction $g_{pred}$ and postprocessing $g_{post}$ are presented in section~\ref{sec:CPM} and section~\ref{sec:BPM}, respectively. 
The architectures of other modules are shown in section~\ref{sec:ID}.

\begin{figure}[t]
    \centering
    \includegraphics[width=1.0\linewidth]{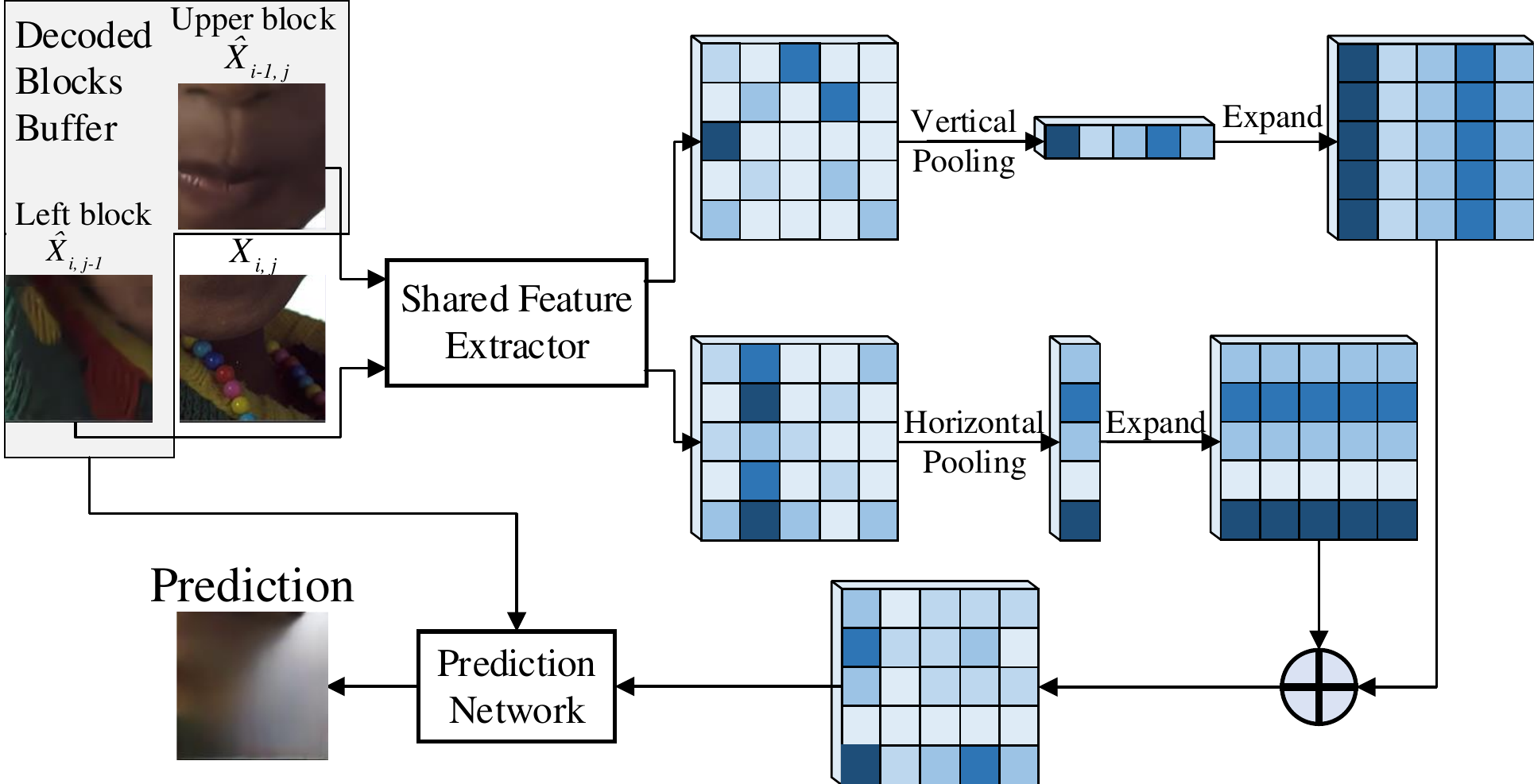}
    \caption{Pipeline of the proposed contextual prediction module (CPM).}
    \label{fig:cpm}
\end{figure}

\subsection{Contextual Prediction Module}
\label{sec:CPM}
In the traditional hybrid coding framework, intra prediction takes previously decoded boundary pixels of the spatially neighboring blocks as context to perform signal prediction~\cite{sullivan2012overview}. However, it only utilizes the boundary information of adjacent blocks, which is not sufficient for learning based image compression.
Going beyond pixels around boundaries, we use all pixels in neighboring blocks to provide long-range contextual information for better prediction.
A simple prediction scheme is to feed the adjacent blocks into a CNN module to predict the current block. However, the module contains a limited receptive field and does not consider the distribution of correlations between pixels of adjacent blocks and the current block. 
To explore a better prediction scheme, we statistically analyze the correlations between pixels in adjacent blocks and current block, which is shown in Fig.~\ref{fig:first}. The correlation between pixels is inversely related to their distance, which means that feeding the reference block directly to the convolution network for prediction is inefficient. 
To capture as much relevant information as possible for prediction, we benefit from the operation of strip pooling~\cite{hou2020strip} and design the contextual prediction module (CPM), which is illustrated in Fig.~\ref{fig:cpm}.

To this end, for enlarging the perceptive field for prediction, we propose to utilize strip pooling to increase the receptive field to cover more relevant pixels in adjacent blocks, thus capturing more correlations from adjacent blocks for prediction. Note that this approach is different from the original use of strip pooling as in Hou et al.~\cite{hou2020strip}, which aims to obtain self-attention information to assist scene parsing. 
\begin{figure}[t]
    \centering
    \includegraphics[width=1.0\linewidth]{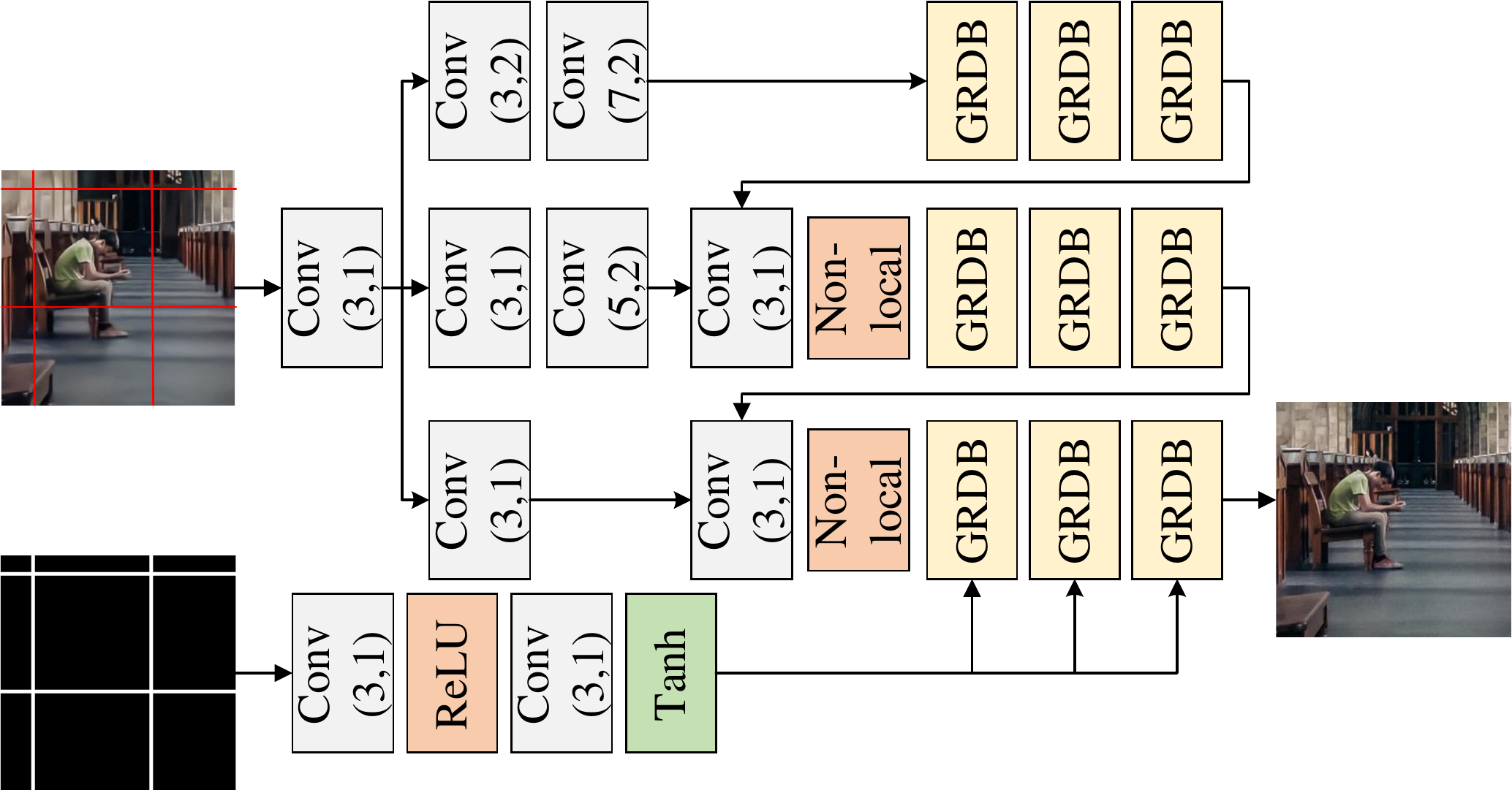}
    \caption{Pipeline of the boundary-aware postprocessing module (BPM). Convolutional parameters are denoted as (kernel size, stride). For information about the detailed nonlocal operations, refer to~\cite{wang2018non}.}.
    \label{fig:bpm}
\end{figure}

Specifically, for the block $\bm{X}_{i, j}$ to be predicted, we utilize the adjacent upper decoded block $\bm{\hat{X}}_{i-1,j}$ and left decoded block $\bm{\hat{X}}_{i,j-1}$ to formulate the prediction result. We first feed them into the shared feature extractor, including convolution operation, normalization and nonlinear activation, to obtain their respective latent representations.
Then, for the feature of the upper block, we utilize the vertical strip pooling method to average feature values in a column, while horizontal strip pooling is used to process the feature of the left block. To fuse the information between the upper and left blocks, 
we extend the pooled features through vertical/horizontal copying and utilize element-wise addition to aggregate them.
After the fusion operation, the fused feature is sent to the prediction network together with the decoded reference block to obtain the prediction. The architecture of the prediction network is based on Unet~\cite{ronneberger2015u}. 
The detailed structure of the feature extractor and prediction network is shown in section~\ref{sec:ID}.

\subsection{Boundary-aware Postprocessing Module}
\label{sec:BPM}
To remove the block artifacts caused by the padding operation on edges, we introduce the concept of the boundary mask to guide the postprocessing module, and further design the boundary-aware postprocessing module (BPM). 
As shown in Fig.~\ref{fig:bpm}, we generate a boundary mask by locating the boundary areas.

We utilize the boundary mask $\bm{M}$ to guide the postprocessing module in two ways. First, we concatenate the coded image and the boundary mask as input and utilize the boundary mask $\bm{M}$ to generate the mask-aware spatial attention to enhance the boundary areas in feature space.
Second, we propose a boundary-aware loss based on the boundary mask. 
The boundary-aware loss can be divided into two parts as shown in Eq (\ref{eq:bl}), namely, the global part $D(\bm{X},\bm{\hat{X}})$ and boundary part $D(\bm{M}\odot \bm{X},\bm{M}\odot \bm{\hat{X}})$.
\begin{equation}
L_{post}=D(\bm{X},\bm{\hat{X}})+\alpha D(\bm{M}\odot \bm{X},\bm{M}\odot \bm{\hat{X}}),
\label{eq:bl}
\end{equation}
where D, $\odot$ and $\alpha$ are the distortion metric, element-wise multiplication and weighting factor, respectively. In this paper, we empirically set the weighting factor to 10. The global part is utilized for the whole image, which can guide the postprocessing module to further enhance the texture and remove global noise. The boundary part is used to enhance the boundary areas of the coded image. 

For the backbone of the postprocessing module, we follow MSGDN~\cite{li2020multi}, which utilizes the multiscale information to remove block artifacts. Here, we utilize the convolution layer with stride 2 to downsample the feature map with $\times$ 1, $\times$ 2, and $\times$ 4, separately. We then utilize the grouped residual dense block (GRDB)~\cite{lu2019learned} to process each feature. The GRDB consists of four residual dense blocks (RDB)~\cite{zhang2018residual}, and each RDB contains a dense layer with 8 convolution layers. To fuse features of different scales, we utilize the convolution with kernel size 1 and nonlocal modules to capture the long dependency in the features. 

\subsection{Block-wise Parallel Processing Strategy}
\begin{figure}[t]
    \centering
    \includegraphics[width=0.55\linewidth]{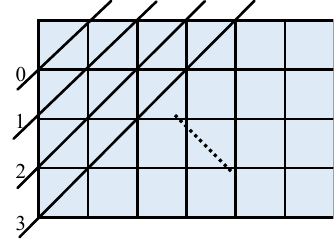}
    \caption{Diagram of parallel acceleration strategy. We process the blocks on the same line in parallel.}
    \label{fig:ac}
\end{figure}

Due to the parallelism of the block-based scheme, we are allowed to compress multiple blocks in parallel to accelerate the coding speed.
Taking into account the characteristics of our prediction scheme, which utilizes the reconstructions of the upper block and left block as reference blocks, we propose a feasible parallelism scheme by setting the coding order of parallel coding blocks as shown in Fig.~\ref{fig:ac}.  Specifically, we utilize a group of $45^{\circ}$ straight lines to determine blocks that can be calculated simultaneously. We number these lines starting from zero and code them from the top line to bottom line. Letting $L$ represent the number of the $L$-th line, where $H_N$ and $W_N$ represent the number of rows and number of columns of the block inside the image, respectively, we can define the set of parallel blocks formed by each line as follows: 
\begin{equation}
S_{L}=\{\bm{X}_{i,L-i}|0\leq i<H_N, 0\leq L-i<W_N\},
\label{eq:set4ac}
\end{equation}
where $S_{L}$ represents the set of blocks that can be calculated in parallel at the $L$-th time. The maximum number of the $L$ is $H_N+W_N-2$ because the last block to be coded is $\bm{X}_{H_{N}-1,W_{N}-1}$. 

For the blocks ${\bm{X}_{0,L}}$, ${\bm{X}_{L,0}}$ at upper and left edges of an coded image, that only have one reference block, we do not perform predictive coding on them. Of cause, training two extra CPM models for above two patterns can further improve the coding efficiency. But the blocks in upper and left edges only take a small part of whole image, while it will increase the network parameters about two times.
On the encoder side, we follow the above scheme to encode the blocks in one frame and obtain the bit-stream. Since the residual decoding does not depend on adjacent decoded blocks, theoretically, we can decode the residuals of all blocks at the same time. Afterwards, we can use the parallel scheme provided in this section to achieve predictive decoding and obtain the final decoded image. 

\begin{figure}[t]
	\centering
	\subfloat[Nonlinear transformation module.]{
	\label{fig:t}
	\includegraphics[width=0.95\linewidth]{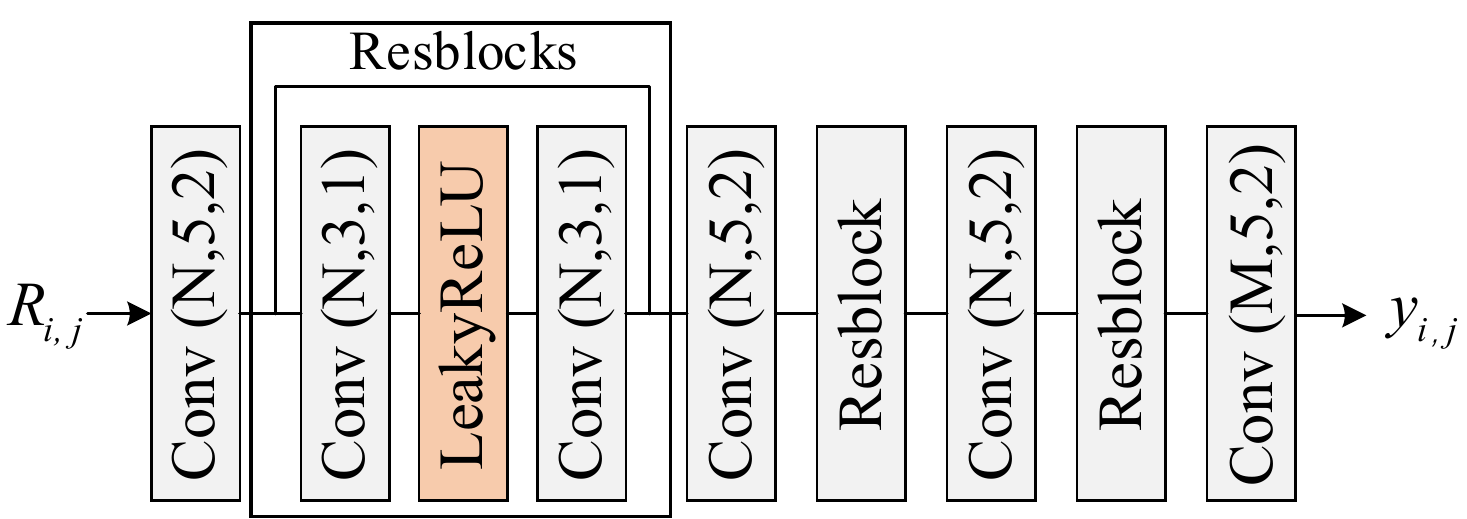}
	}\\
	\subfloat[Inverse nonlinear transformation module.]{
	\label{fig:it}
	\includegraphics[width=0.85\linewidth]{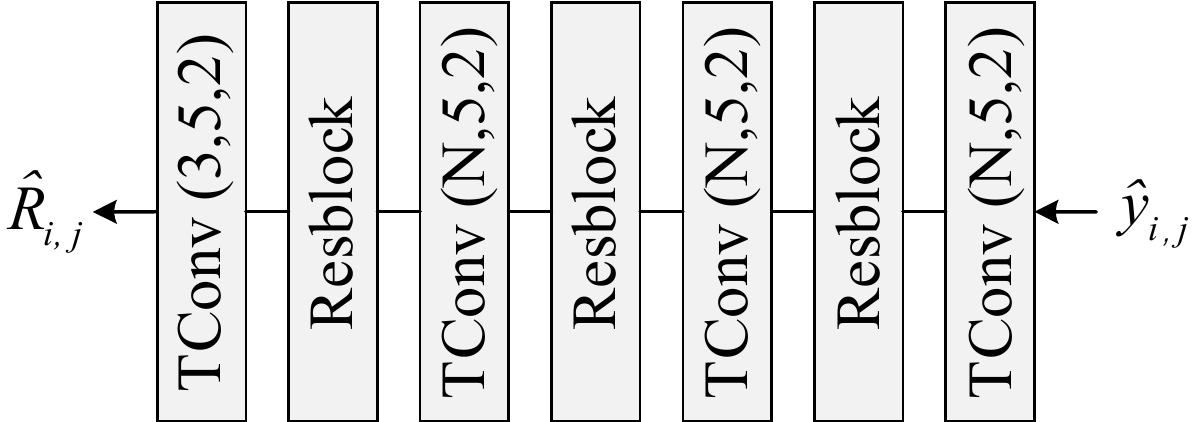}
	}\\
	\subfloat[Entropy model.]{
	\label{fig:entropy}
	\includegraphics[width=1.0\linewidth]{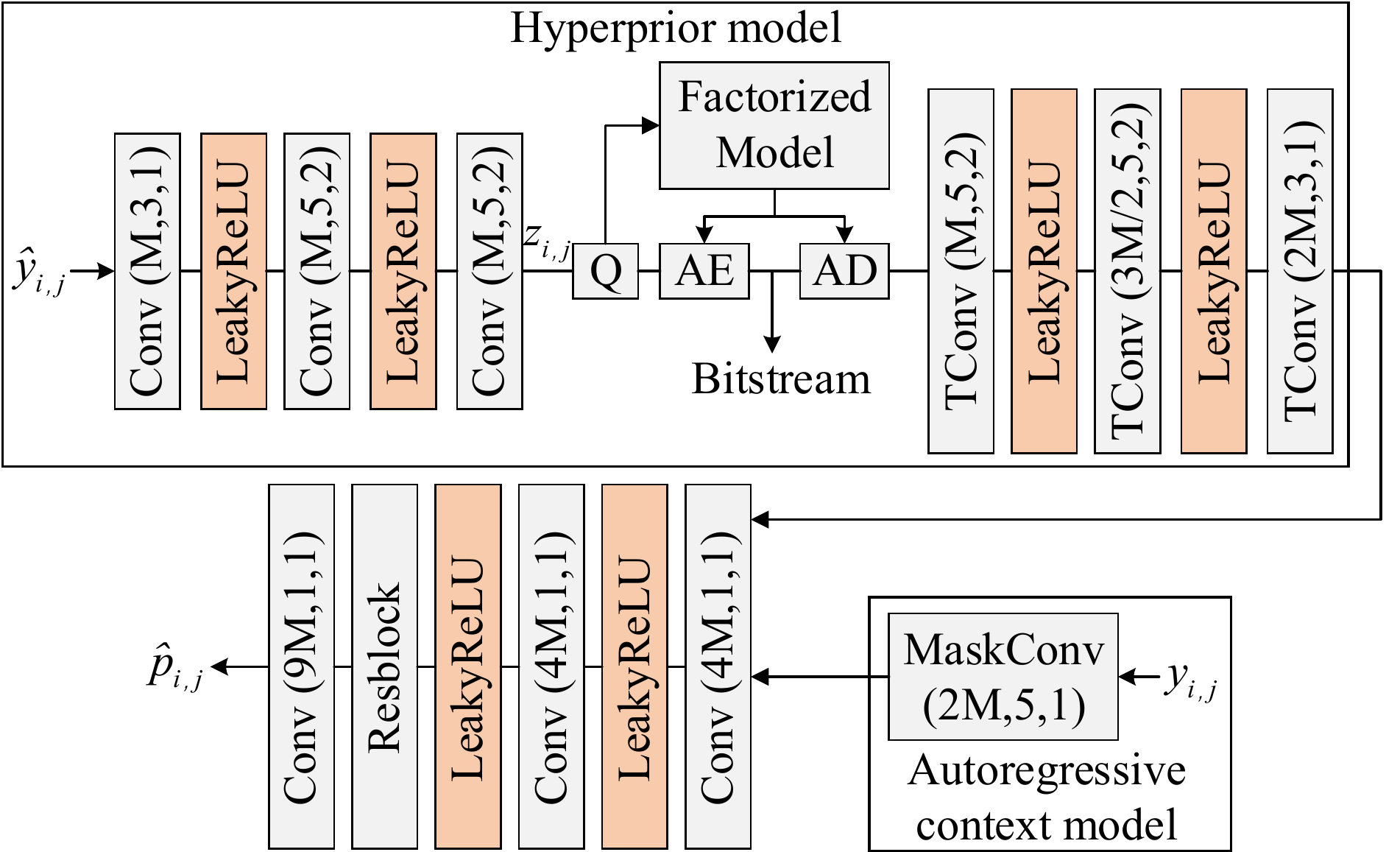}
	}\\
	\caption{Detailed layer information in LBHIC. Q, AE and AD denote quantization, arithmetic encoder, and arithmetic decoder, respectively. Conv represents the convolution operation, and TConv stands for the transpose convolution operation. The parameters of Conv and TConv are denoted as (number of filters, kernel size, stride). The parameters N and M are selected according to the corresponding rate-distortion model.}
	\label{fig:moduleincoder}
\end{figure}


\subsection{Implementation Details}
\label{sec:ID}

\paragraph {Quantization} We utilize the round operation as the quantization $Q$ in our LBHIC. Since the round operation is nondifferentiable in neural training, we follow Ball\'{e} et al.~\cite{balle2017end} to utilize additive uniform noise $u$ as the substitution in the training stage:
\begin{equation}
\bm{\hat{y}}_{i, j} = Q(\bm{y}_{i, j})=\bm{y}_{i, j}+u(-\frac{1}{2},\frac{1}{2}).
\label{eq:sub1}
\end{equation}
However, the additive noise will cause a mismatch between the training and testing stage. To solve this problem, we straight-through the gradients to substitute gradients of the round operation, which can be formulated as:
\begin{equation}
\bm{\hat{y}}_{i, j} = Q(\bm{y}_{i, j})=\bm{y}_{i, j} - Stop(\bm{y}_{i,j} - \lfloor \bm{y}_{i,j} \rceil),
\label{eq:sub2}
\end{equation}
where $\lfloor \rceil$ denotes the round operation and $Stop$ denotes stopping the gradient. Note that we only utilize the approximation of the Eq (\ref{eq:sub2}) in the input of the inverse nonlinear transformation, while the input of the entropy model still adopts the quantization method in Eq (\ref{eq:sub1}).

\begin{figure}[t]
	\centering
	\subfloat[Structure of feature extractor.]{
	\label{fig:fe}
	\includegraphics[width=0.8\linewidth]{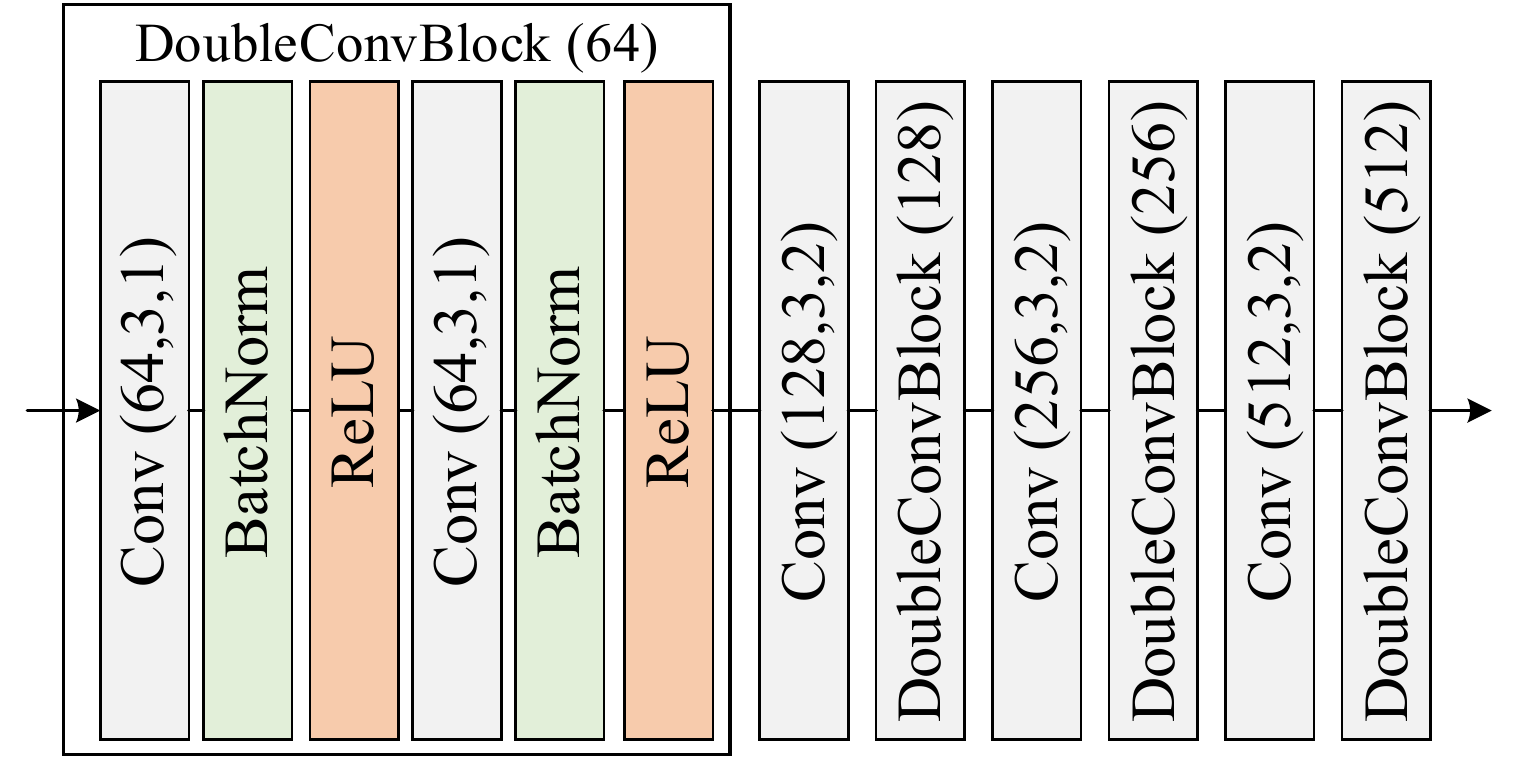}
	}\\
	\subfloat[Structure of prediction network.]{
	\label{fig:pn}
	\includegraphics[width=1.0\linewidth]{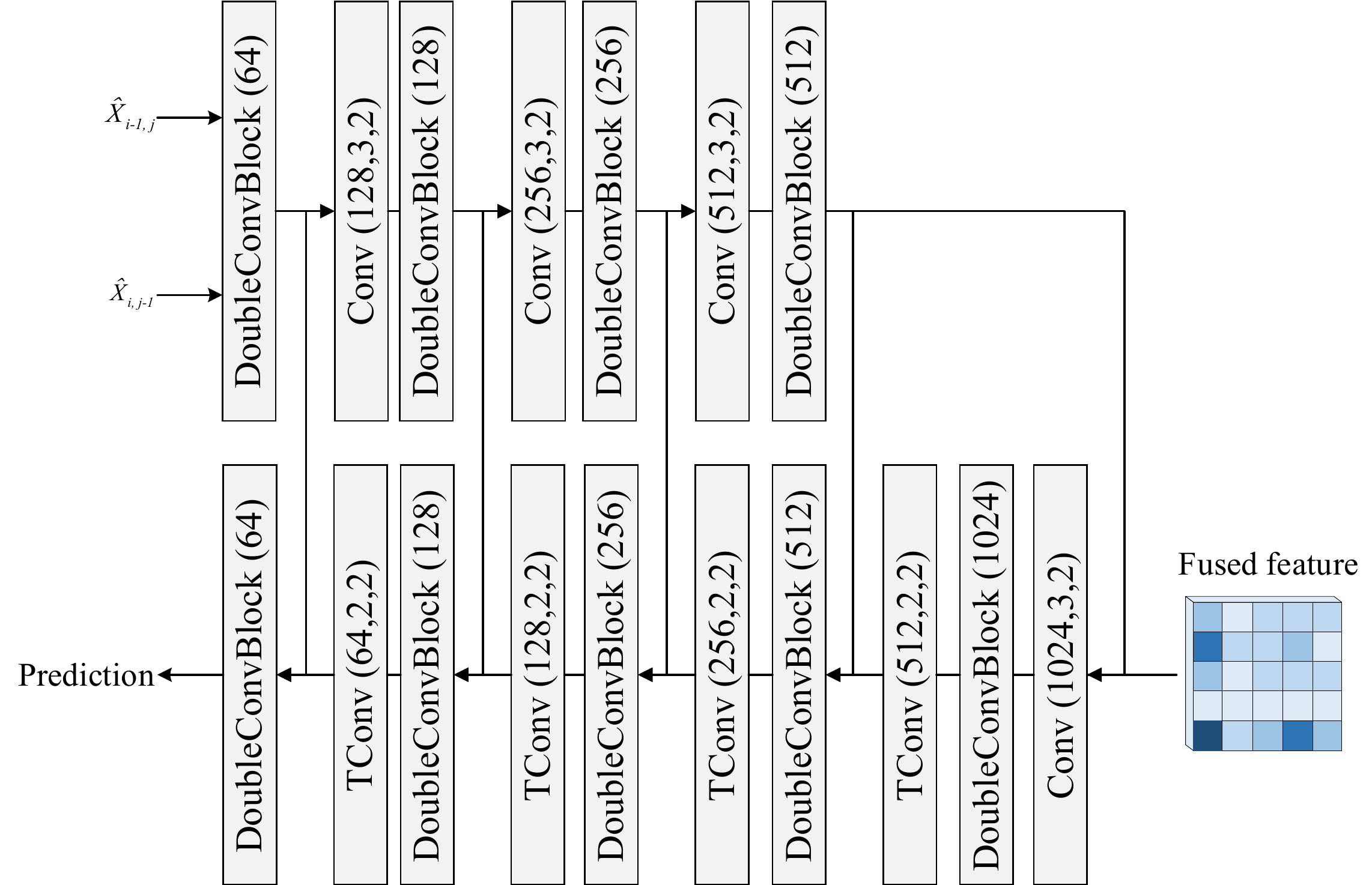}
	}\\
	\caption{Structures of feature extractor and prediction network in CPM.}
	\label{fig:moduleinCPM}
\end{figure}


\paragraph {Model Training} We jointly train all modules of LBHIC, except the BPM, through rate-distortion optimization. Specifically, the rate and distortion are weighted against each other with a Lagrange multiplier $\lambda$ as:
\begin{equation}
L = rate + \lambda Distortion,
\label{eq:rdo}
\end{equation}
where the distortion term can be any differentiable metric in our framework. According to information theory, the minimum code rate is the cross entropy of the real distribution $\bm{p}_{i,j}$ and the estimated distribution $\bm{\hat{p}}_{i,j}$, which can be calculated by the following formula:
\begin{equation}
rate = E_{\bm{p}_{i,j}}[-log(\bm{\hat{p}}_{i,j})].
\label{eq:rate}
\end{equation}

In the experiments, we train different models with different $\lambda$ to evaluate the rate-distortion performance for various ranges of bit-rate.
For MSE, we evaluate our models with seven $\lambda$ values: 128, 256, 512, 1024, 2048, 4096 and 8192. For MS-SSIM, we evaluate the rate-distortion performance of our LBHIC with six $\lambda$ values: 8, 16, 32, 64, 128 and 256.  
The network training includes four stages. First, we train the baseline model, that is, the structure of LBHIC without CPM and BPM. The number $N$ of the GMM model is set to 3 in our baseline model. Second, we train CPM to predict the block according to the decoded reference blocks. Then, we jointly train the baseline model with CPM in an end-to-end manner. Finally, we train BPM to remove the block effects caused by the block-based schemes.
In all stages, we use Adam~\cite{kingma2014adam} as the optimizer to train the network, and the initial learning rate is set to $5\times 10^{-5}$. For the third stage, we increase the learning rate decay by 0.5 after every 300,000 iterations. The minibatch size is set as 8, and the whole system is implemented based on PyTorch.

\paragraph {Detailed Layer Information} In this section, we provide a detailed layer description of the nonlinear transformation, inverse nonlinear transformation, and entropy model, as shown in Fig.~\ref{fig:moduleincoder}. 
We adapt N as 128 and M as 192 in five low-rate models. For the other high-rate models, N and M are  set as 256 and 448, respectively, which requires more channels to preserve high-frequency information. In the entropy model, the hyperprior model~\cite{balle2018variational} and the autoregressive context model~\cite{minnen2018joint} are both utilized to estimate the probability of elements in the feature $\bm{\hat{y}}_{i,j}$.  We also provide the detailed structures of the feature extractor and prediction network in our CPM, which are illustrated in Fig.~\ref{fig:moduleinCPM}. In BPM, GRDB is our basic unit. For the basic architecture of the GRDB, refer to \cite{kim2019grdn}. Here, we set the RDB number in each GRDB to 4 and the number of layers in each RDB to 8.

\paragraph{Parallel Threads} In this paper, the maximum number of the parallel threads we set is 8. As shown in Table \ref{tab:timebd}, the encoding time is 0.8780s and decoding time is 4.9419s with the kodak dataset. When the number of parallel thread is set 1, the encoding time is 6.8026s and decoding time is 24.1270s, which can evaluate the effectiveness of our proposed parallel scheme. 
\begin{figure}[t]
	\centering
	\subfloat[PSNR result on Kodak dataset.]{
	\label{fig:exppsnr}
	\includegraphics[width=0.7\linewidth]{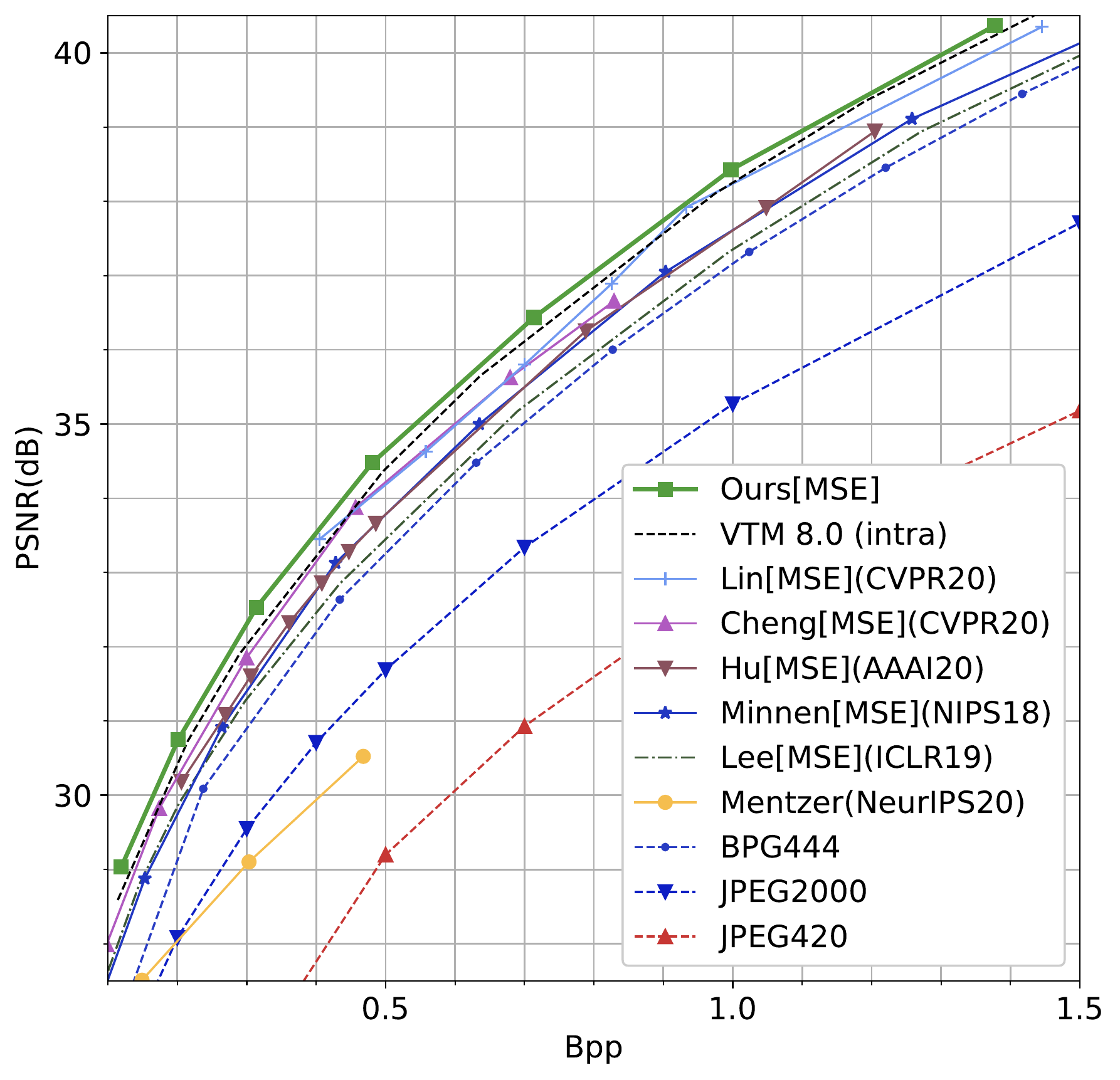}
	}\\
	\subfloat[MS-SSIM result on Kodak dataset.]{
	\label{fig:expssim}
	\includegraphics[width=0.7\linewidth]{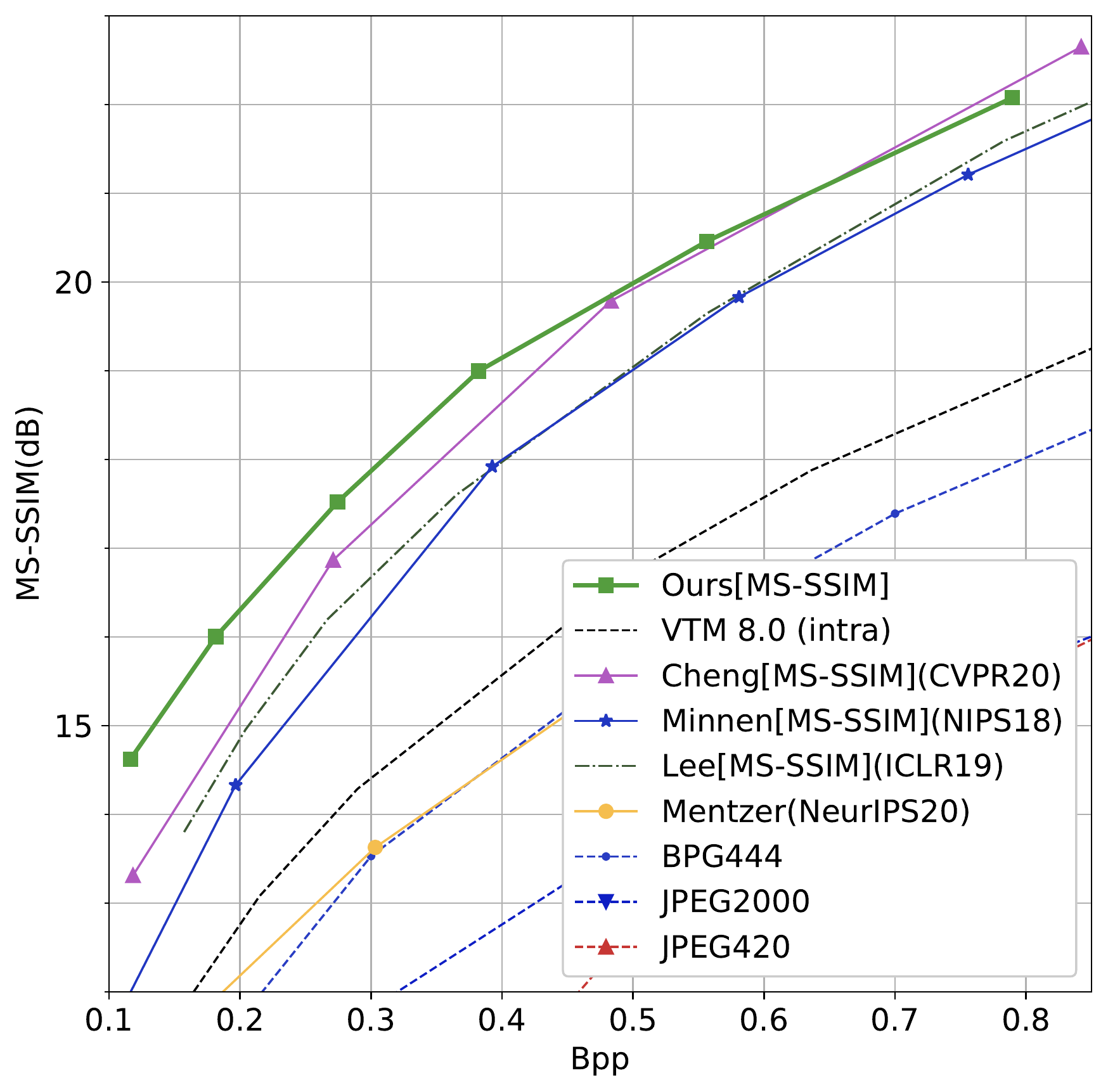}
	}\\
	\subfloat[PSNR result on Tecnick dataset.]{
	\label{fig:tecperformance}
	\includegraphics[width=0.7\linewidth]{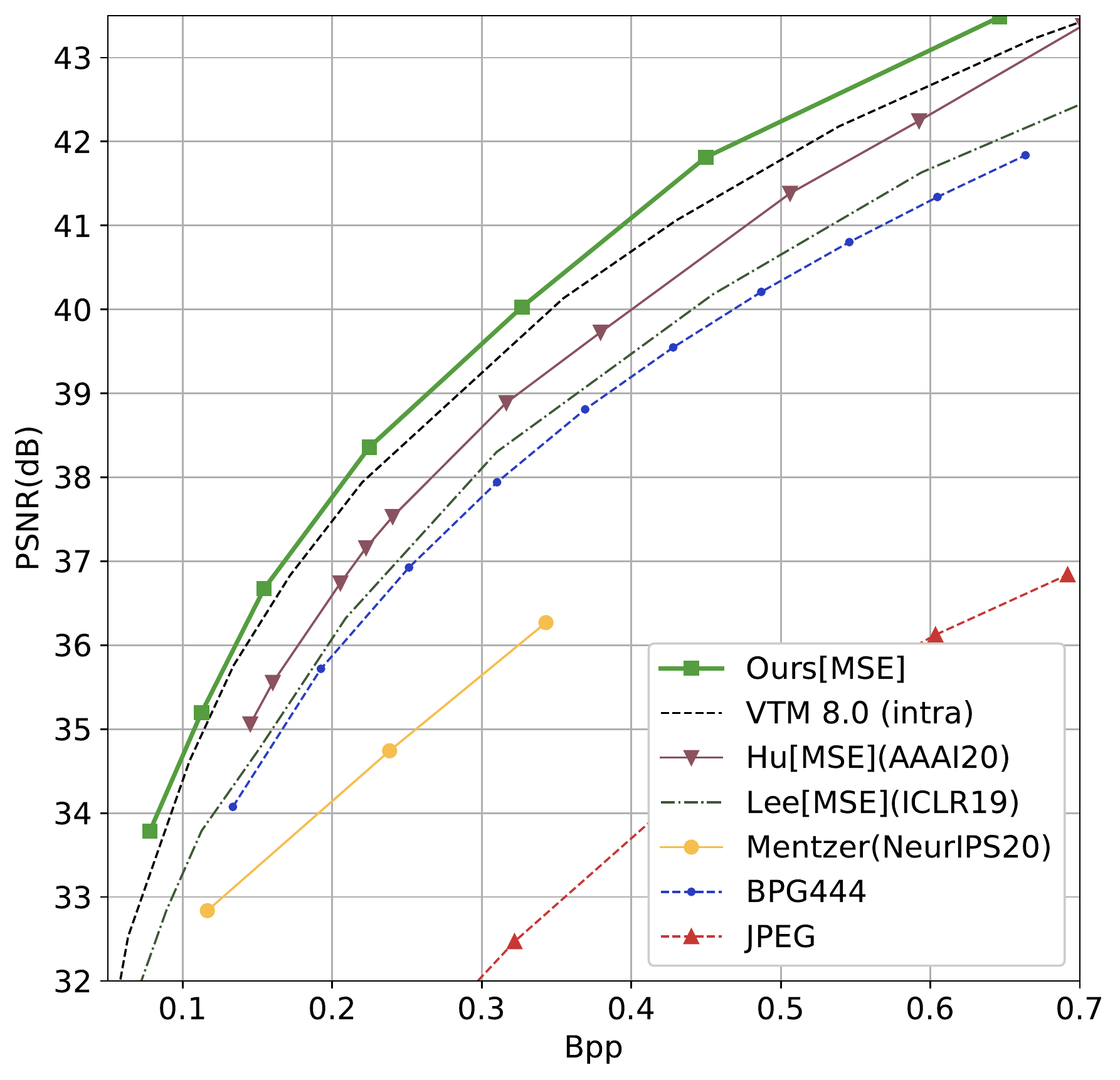}
	}\\
	\caption{R-D performance covering a wide range of traditional and learned compression methods.}
	\label{fig:performance}
\end{figure}

    


\begin{figure*}[t]
\setlength{\abovecaptionskip}{0pt} 
\setlength{\belowcaptionskip}{-0pt}
    \centering
    \includegraphics[width=0.88\linewidth]{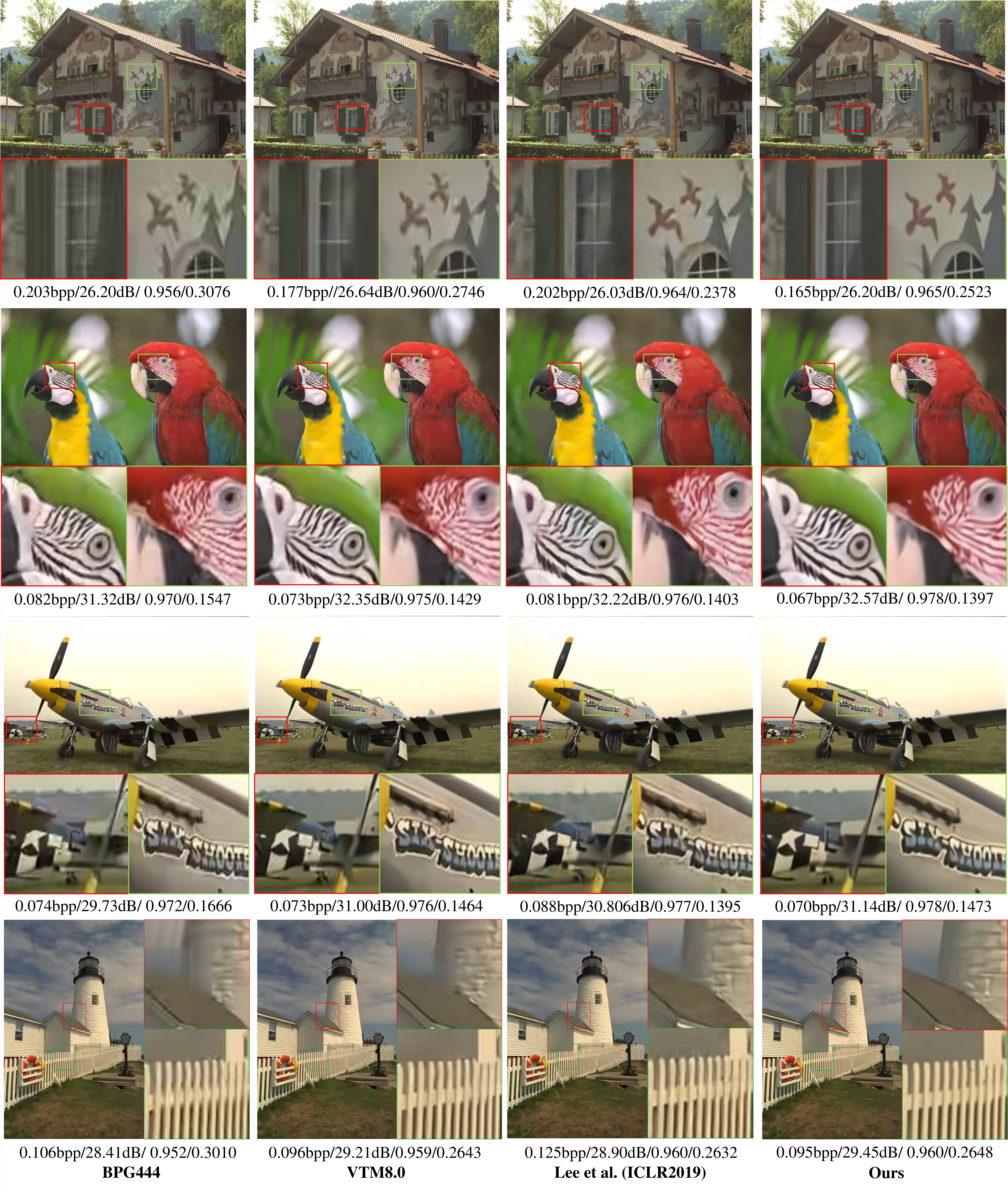}
    \centering
    \caption{Comparisons of BPG, VTM8.0 (intra), Lee \cite{Lee2019Context} and our LBHIC with low rate  in terms of BPP/PSNR/SSIM/LPIPS.}
    \label{fig:sub_results}
\end{figure*}

\section{Experiment Results}
\label{sec:4}
\subsection{Experimental Setup}
\paragraph {Datasets}
We use Imagenet~\cite{deng2009imagenet} as the training set for the first three training stages, and utilize the DIV2K dataset~\cite{agustsson2017ntire} and CLIC 2020 challenge training set \href{https://challenge.compression.cc/tasks} for the fourth training stage. We augment training datasets by randomly cropping the images into $256\times256$ patches. To evaluate the performance of our method, we utilize the Kodak dataset \cite{franzen1999kodak} and Tecnick SAMPLING dataset \cite{asuni2014testimages} as our test dataset. The Kodak dataset contains 24 lossless images of size $512 \times 768$, while the Tecnick dataset contains 40 images of size $1200 \times 1200$.

\paragraph {Evaluation Setting} 
We measure the quality of reconstructed images with the metrics PSNR and MS-SSIM~\cite{wang2003multiscale}, and we calculate bits per pixel (bpp) for each image to measure coding cost. To verify the effectiveness of our LBHIC, we compare our LBHIC with the state-of-the-art learned image compression methods~\cite{lin2020spatial, cheng2020learned, hu2020coarse, balle2018variational, minnen2018joint} and the traditional codecs (JPEG, JPEG2000, BPG, and VVC). In this paper, we evaluate VVC by using VVC standard reference software VTM 8.0. 
Specifically, we use the VTM $encoder\_intra\_vtm.cfg$ as our configuration file, and we set InputChromaFormat=444 for all datasets. Given an input image in RGB format, we first convert it to YUV444 format and then compress it with VTM 8.0 (intra). Afterwards, the reconstructed file is converted back into RGB color space for evaluation.
We also compare the encoding and decoding speeds for different learning-based codecs. To ensure fairness, we test these schemes on the same machine (Intel Core i7-8700 CPU / NVIDIA GTX 1080 GPU).

\begin{table*}[t]
\caption{Quantitative evaluation results compared with recent works.}
\centering
\begin{tabular}{lccccccc}
\toprule
\multicolumn{1}{c}{Dataset} & Method         & Encode  time (s) & Decode time (s) & GPU VRAM(MiB) & BD-rate (\%) & \begin{tabular}[c]{@{}c@{}}Low rate \\ model size (MB)\end{tabular}  & \begin{tabular}[c]{@{}c@{}}High rate \\ model size (MB)\end{tabular} \\ \midrule
\multirow{5}{*}{Kodak}      & VTM (Anchor) & 74.9985  &  0.1074        &--               & 0.0     & 7.2   & 7.2     \\
                            & Lee (ICLR19)   & 10.7854          & 27.7725  & 1173                & 17.0  & 123.8 &  292.6    \\
                            & Hu (AAAI20)    & 13.2832          & 38.2971  & 1437                 & 11.1  & 84.6 & 290.9       \\
                            & Cheng (CVPR20) & 24.3037          & 19.6567  & 1387            & 4.8   & 134.3 & 299.11   \\
                            & Ours           & 0.8780           & 4.9419   & 1247               & -4.1  & 288.8 & 474.6    \\ \midrule
\multirow{5}{*}{Tecnick}    & VTM (Anchor)& 165.9363   &   0.2624       & --               & 0.0   & 7.2 & 7.2       \\
                            & Lee (ICLR19)   & 38.617672        & 101.5829 & 2945                 & 30.3   & 123.8 &  292.6       \\
                            & Hu (AAAI20)    & 29.4867          & 116.8766 & 3459            & 16.8  & 84.6 & 290.9   \\
                            & Cheng (CVPR20) & 64.4792          & 62.9775  & 5009            & 6.5   & 134.3 & 299.11       \\
                            & Ours           & 3.1255           & 16.7236  & 1247                 & -6.4   & 288.8 & 474.6      \\ \bottomrule
\end{tabular}
\label{tab:timebd}
\end{table*}

\subsection{Overall Performance Results}
As shown in Fig.~\ref{fig:exppsnr} and Fig.~\ref{fig:expssim}, we visualize RD curves of our LBHIC model and the SOTA methods (including Lin (CVPR20)\cite{lin2020spatial}, Cheng (CVPR20)\cite{cheng2020learned}, Hu (AAAI2020)\cite{hu2020coarse}, Minnen (NIPS18)\cite{minnen2018joint}, Mentzer (NIPS2020)\cite{mentzer2020high} and Lee (ICLR19)\cite{Lee2019Context}) in terms of the PSNR and MS-SSIM metrics on the Kodak dataset, respectively. Our methods outperform all previous methods with respect to both metrics with the same coding cost. We also compare our LBHIC with traditional codecs (including JPEG, JPEG2000, BPG, and VTM8.0). Compared with the latest traditional codec VTM 8.0 (intra), our LBHIC framework can achieve approximately 0.3 dB gain in PSNR at all rate points. Since the learned codec can flexibly learn better representation for a given distortion metric, our method achieves better MS-SSIM performance than the traditional codec VTM8.0. In addition, owing to the CPM and BPM, our model achieves obvious improvements compared with the learned codecs. To further verify the generalization performance of our LBHIC, we directly apply our LBHIC to encode the Tecnick dataset with the same weights utilized to encode the Kodak dataset. The corresponding performance is shown in Fig.~\ref{fig:tecperformance}. Our model can be generalized to high-resolution (1200$\times$1200) image datasets and can still outperform existing traditional (VTM 8.0, BPG, and JPEG) and learned image compression methods~\cite{hu2020coarse,Lee2019Context} in terms of PSNR.

We also provide the subjective comparison of our LBHIC framework with traditional codecs (including BPG444 and VTM 8.0 (intra)) in the low rate setting. Fig. \ref{fig:sub_results} shows the raw image and the results produced by the three compression methods. Visual artifacts such as blur, ringing, and block effects can be observed in reconstructions of the traditional codecs (such as the graffiti on the wall on the first line, the text on the airplane on the third line, and the roof on the last line).
Compared with BPG444 and VTM 8.0 (intra), our scheme eliminates the ringing effect, noising and blurring under similar rate cost, providing better visual quality. 

\begin{table}[b]
\caption{Prediction performance of CPM compared with the SOTA prediction methods on the Kodak dataset.}
\centering
\begin{tabular}{ccccc}
\toprule
 & \begin{tabular}[c]{@{}c@{}}Prediction \\ network\end{tabular} & \begin{tabular}[c]{@{}c@{}} Dumas et al. \\ (TIP19) \end{tabular}& \begin{tabular}[c]{@{}c@{}} Hu et al. \\ (TMM19)\end{tabular} & CPM\\
\midrule
\begin{tabular}[c]{@{}c@{}}PSNR (dB) of \\ Prediction \end{tabular} & 16.6812 & 18.9015 & 19.0539 & 19.4279 \\ \midrule
\begin{tabular}[c]{@{}c@{}}PSNR (dB) of \\ Reconstruction \end{tabular} & 30.2551 & 30.2882 & 30.3795 & 30.4426 \\ \midrule
\begin{tabular}[c]{@{}c@{}}MS-SSIM of \\ Reconstruction \end{tabular} & 0.9747 & 0.9749 & 0.9755 & 0.9758 \\ \midrule
Bpp & 0.2002 & 0.2008 & 0.2034 & 0.2012\\\midrule
Params (MB) & 106.01 & 76.52 & 960.59 & 196.05 \\ \midrule
\begin{tabular}[c]{@{}c@{}} Average \\ Runtime (ms)\end{tabular} & 12.82 & 7.15 & 30.60& 21.99 \\ \bottomrule
\end{tabular}
\label{tab:prediction}
\end{table}

To analyze the encoding and decoding time complexity of our codec, in Table~\ref{tab:timebd}, we first reproduce SOTA learned image compression methods (including~\cite{lee2019hybrid,hu2020coarse,cheng2020learned}) and compare the encoding and decoding runtime on the same machine to ensure fairness. The encoding and decoding runtime are measured by averaging the encoding and decoding time across all images in Kodak dataset. Then, we calculate the BD-rate~\cite{bjontegaard2001calculation} of each method with the anchor VTM 8.0 (intra) 
, where the BD-rate metric represents the average rate saving with equivalent distortion between two methods. Here, we utilize the PSNR as the distortion metric.
Compared with existing learning based solutions, our method provides better compression performance while providing nearly 5x improvement in decoded runtime (i.e., a decoding time savings of 74.8\% compared with that in Cheng~\cite{cheng2020learned}) on Kodak dataset. Under the acceleration of GPU, our decoding time is 4.9419s.  Without GPU acceleration, our decoding time is 62.9160s. 
It is worth noting that even compared with the latest traditional standard VVC (VTM 8.0), our solution can save 4.1\% of bits while maintaining the same compression quality. On Tecnick dataset, our method also provide similarly improvement in decoded runtime, and achieves 6.1\% bit saving compared with VVC. Although our scheme still has a gap in runtime compared with traditional codec, we think that the proposed method provides a feasible direction for acceleration on the learning based scheme, which provides enough potential for practice application. We believe that with the development of GPU chip technology and the further optimization of engineering, it can achieve preformance exceeding in all aspects compared with traditional codec. To demonstrate the friendliness of our solution to GPU memory, we give the GPU memory consumption of different solutions in Table~\ref{tab:timebd}. We can see that the GPU consumption of existing solutions will increase with the increase of image resolution. Since the coding unit of our solution is a fixed-size block, its video memory usage will not increase with the increase in resolution, which is more hardware-friendly. In the last two columns of the Table I, we compare the model sizes of different methods. It can be seen that due to the addition of the prediction module, our scheme has nearly doubled the model size compared with existing methods. But we believe that this increase in storage is worthwhile in terms of the nature of fixed GPU memory and the benefits of acceleration.

\subsection{Ablation Study}
\subsubsection{Effectiveness of the Contextual Prediction Module (CPM)}
Since there are fewer works which attempt to utilize explicit prediction in learned image compression, here, we compare our CPM with some neural network based prediction methods~\cite{dumas2019context, hu2019progressive} that have been proposed for the traditional hybrid coding framework. Specifically, we reproduce their works and integrate them into our learned compression framework. We then utilize mean square error as the goal of distortion optimization and train each model with the same rate-distortion point ($\lambda=256$), which is trained on the same training set (Imagenet~\cite{deng2009imagenet}). 
To measure the quality of prediction, we 
calculate the distortion between original blocks and predicted blocks for each of the prediction methods. Besides, we further analyze the model parameters and time complexity of each model. The qualitative prediction results on the Kodak dataset are shown in the first row of Table~\ref{tab:prediction}. Compared with existing prediction methods (including Dumas et al.~\cite{dumas2019context} and Hu et al. \cite{hu2019progressive}), our CPM can achieve better prediction quality with a higher PSNR. Although compared with Prediction network and Dumas et al., our CPM adds additional model parameters and runtime, the runtime of the prediction actually accounts for a minor part of the coding time. Hence, it is worth paying for the above minor overhead for performance.

We also attempt to directly include the pixel row/column used for prediction in traditional codec into our prediction network CPM. And then, we retrain the prediction network from scratch. The prediction result is 19.4358dB, which is only a gain of 0.008dB compared with our CPM and can be ignored. 

\begin{figure}[t]
    \centering
    \includegraphics[width=1.0\linewidth]{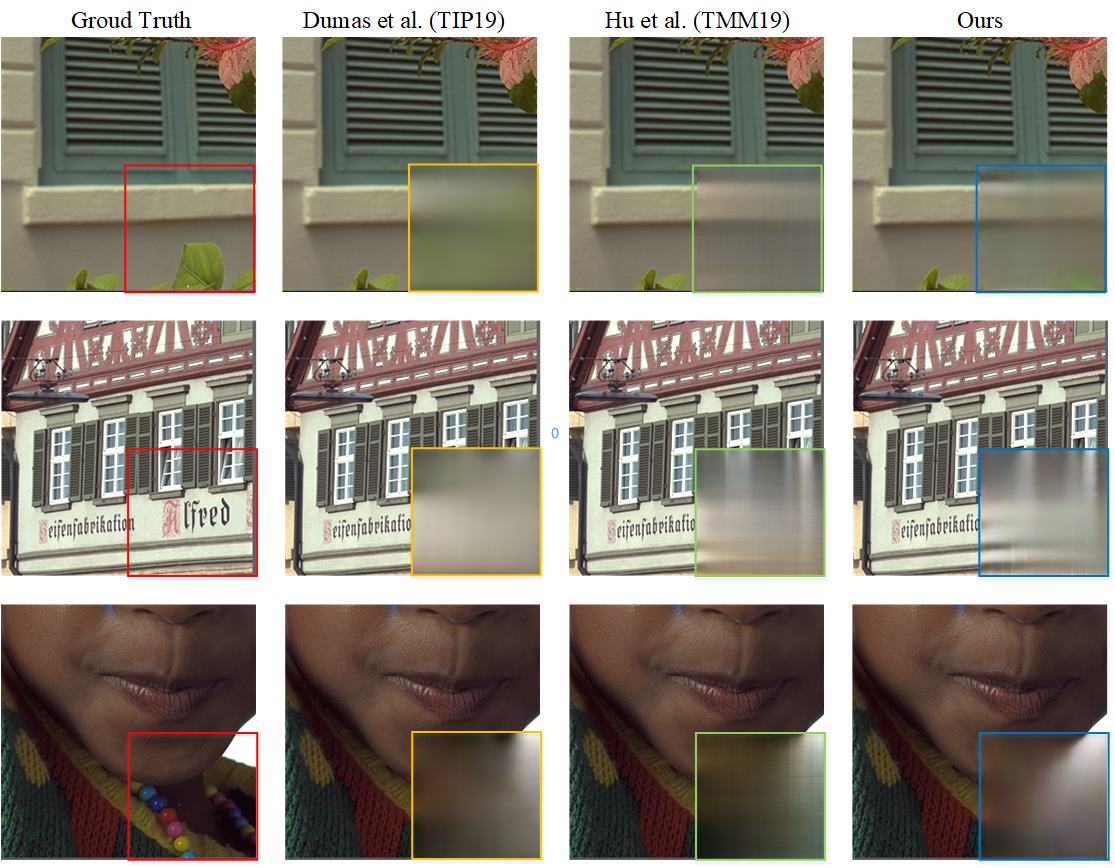}
    \caption{Visualization of the prediction result.}
    \label{fig:vscpm}
\end{figure}

To analyze the prediction results more intuitively, we visualize results between our CPM and the above existing prediction results, as seen in Fig.~\ref{fig:vscpm}. 

\begin{figure}[b]
    \centering
    \includegraphics[width=0.9\linewidth]{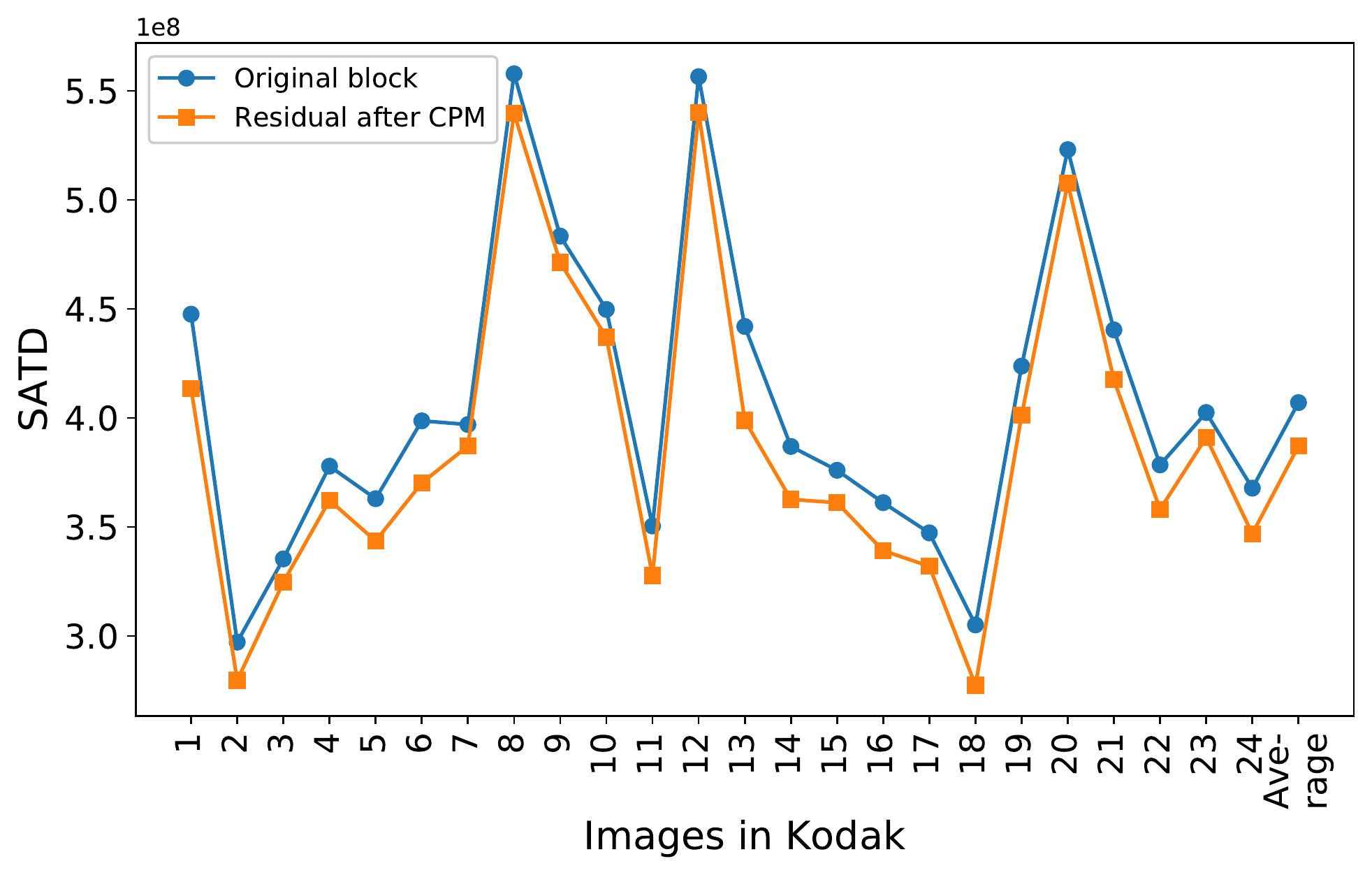}
    \caption{SATD of the original image and corresponding residual on the Kodak dataset.}
    \label{fig:SATD}
\end{figure}

From Fig.10, we can see that our prediction scheme can predict continuous textures (window edges in row 1, wall pattern in row 2, face profile in row 3, etc.) through context blocks. For the contents that cannot be provided by neighboring blocks, they cannot be predicted completely (leaves in row 1, text in row 2, necklace in row3, etc.) and they can be further reconstructed by residual coding.
We also provide corresponding rate and distortion evaluation of reconstructed blocks in the second and third rows of Table~\ref{tab:prediction}, which proves the effectiveness of our CPM in improving compression performance.

\begin{table}[b]
\caption{Performance of BPM compared with the SOTA postprocessing methods on the Kodak dataset.}
\centering
\begin{tabular}{ccccc}
\toprule
Method & DHDN & GRDN & CLIC19\_MS & BPM \\ \midrule
PSNR (dB) & 30.617 & 30.614 & 30.488 & 30.702 \\
MS-SSIM  & 0.9766 & 0.9766 & 0.9761 & 0.9769 \\
Params (MB) & 168.2  & 112.43 & 6.36 & 50.30 \\
Runtime (s) &  0.0710 & 0.0340  &  0.0086 & 0.0936\\\bottomrule

\end{tabular}
\label{tab:post}
\end{table}

The goal of the prediction is to predict as accurately as possible, thereby reducing the transmitted rate of the residual to be compressed. Therefore, we design an experiment to indirectly verify the effectiveness of our CPM by estimating the rate of our residual. Specifically,
we utilize the sum of absolute transformed differences (SATD)~\cite{richardson2004h} to measure the rate size required to transmit the information. We first transform the input through the Hadamard transformation and then calculate the sum of absolute transformed coefficients to obtain the SATD. We calculate the SATD of the original images on the Kodak dataset and calculate the SATD of the corresponding residual obtained by the low rate model ($\lambda=256$), and the result is shown in Fig.~\ref{fig:SATD}. Compared with the original image, the residual information obtained after prediction has a smaller SATD on all test images, which means that our CPM removes the correlation between blocks and obtains more compressible residual representation. 

\begin{figure}[t]
    \centering
    \includegraphics[width=1.0\linewidth]{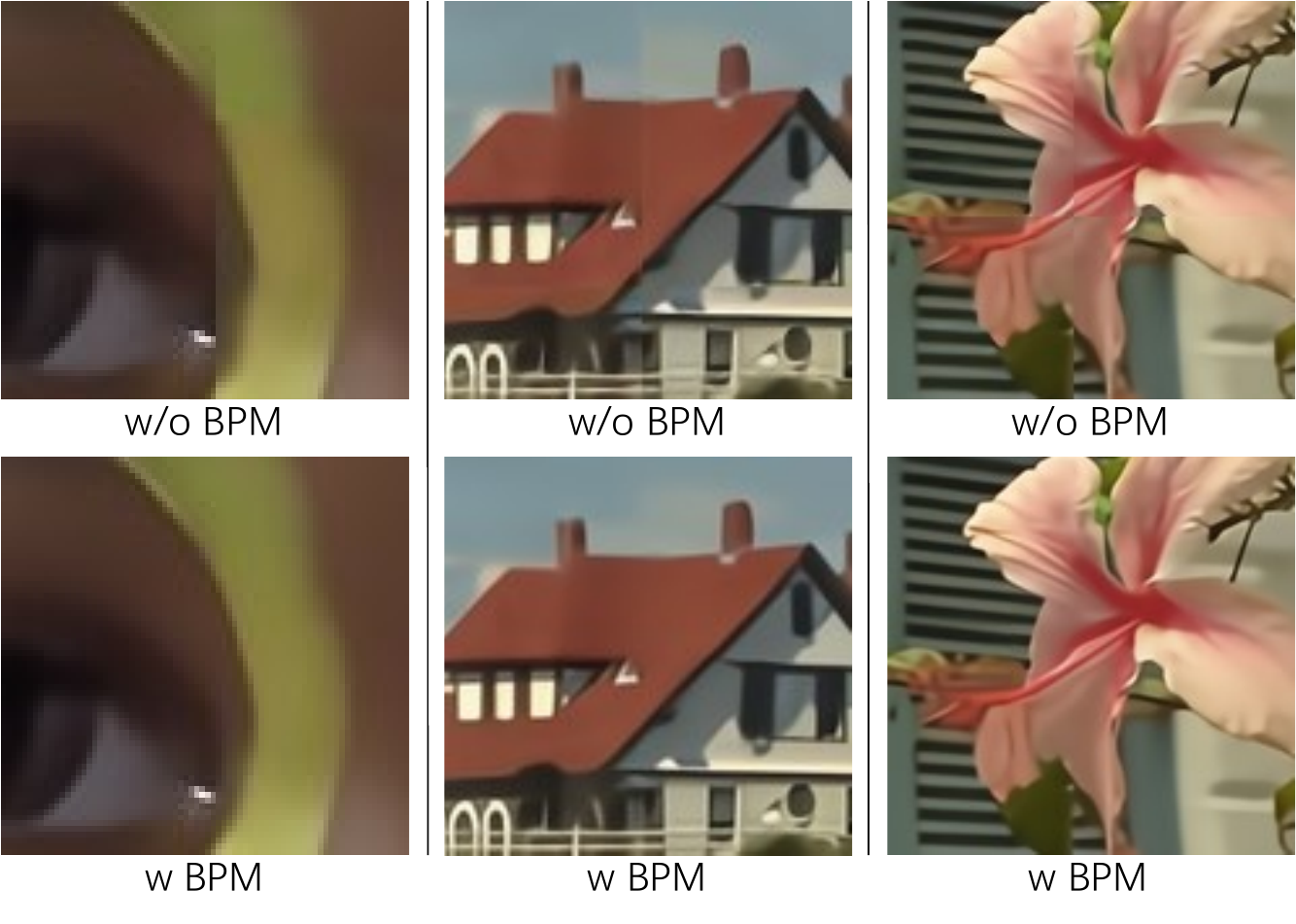}
    \caption{Visualization of image qualities before and after the boundary-aware postprocessing module (BPM).}
    \label{fig:vsbpm}
\end{figure}

\begin{figure}[t]
	\centering
	\subfloat[Contribution of the proposed modules.]{
	\label{fig:ab1}
	\includegraphics[width=0.9\linewidth]{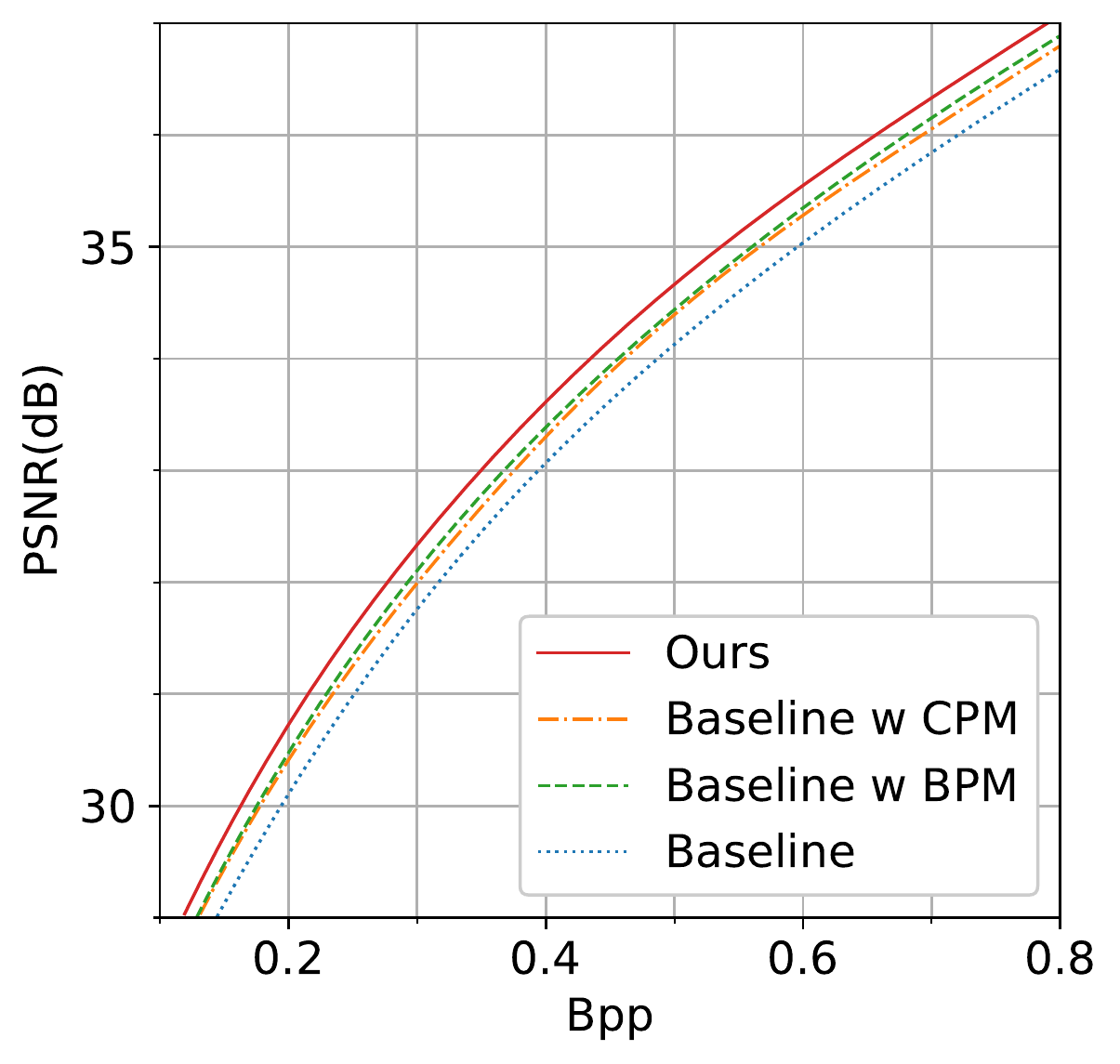}
	}\\
	\subfloat[Influence of block size.]{
	\label{fig:ab2}
	\includegraphics[width=0.85\linewidth]{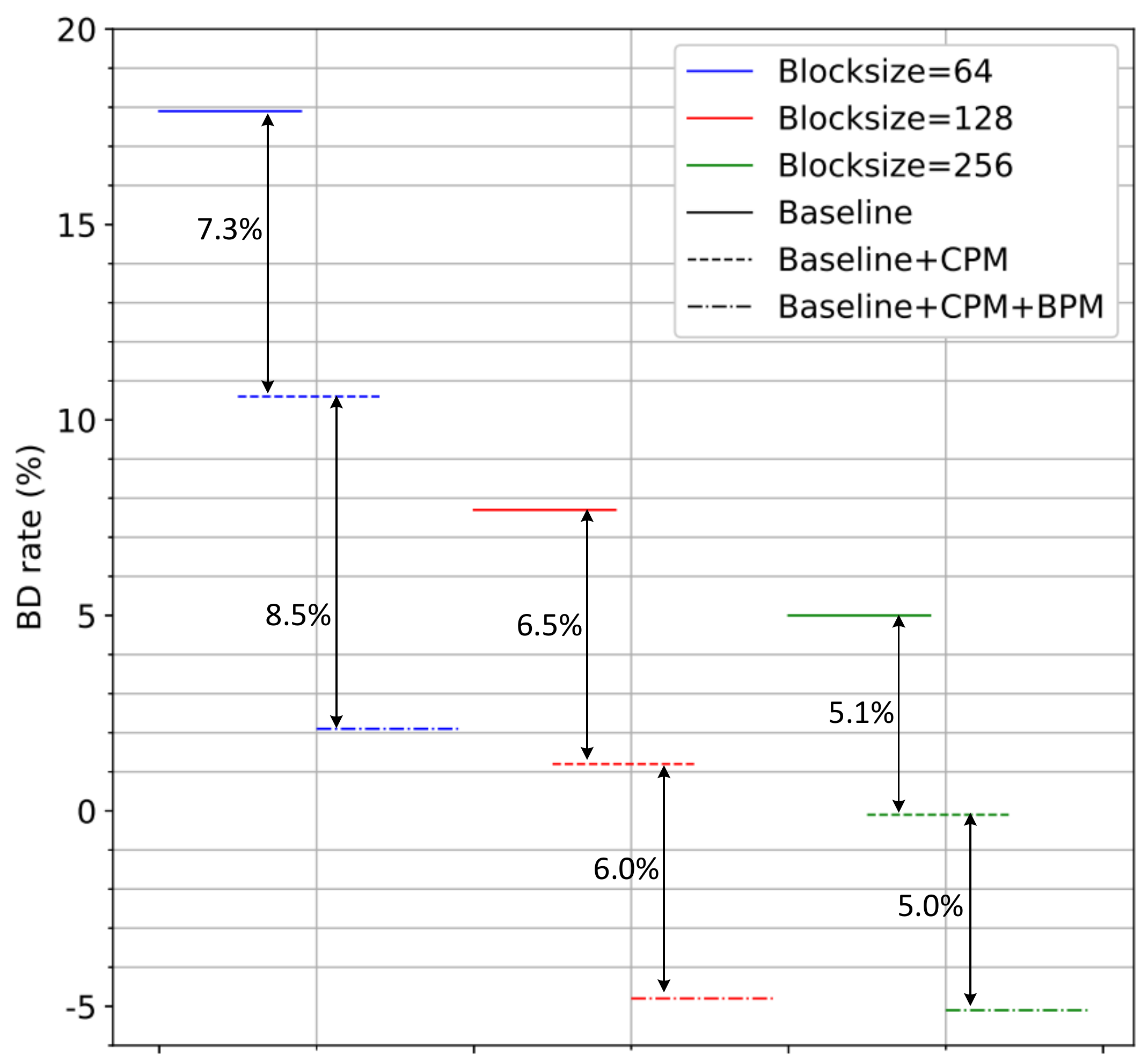}
	}\\
	\caption{Ablation study.}
	\label{fig:ab}
\end{figure}


\subsubsection{Effectiveness of the strip operation}
To prove the effectiveness of strip operation in our CPM, we first simply concatenate reference blocks. Specifically, for the reference block feature with a feature shape of $H\times W\times 3$, the operation of concatenation is to connect the above two features in the channel(RGB) dimension. After the concatenation, we feed them into the prediction network (Fig.~\ref{fig:cpm}) to test its prediction results, which is labeled as prediction network in Table~\ref{tab:prediction}. Without the strip operation, the prediction network can achieve only approximately 16.6812 dB for prediction, which is 2.7 dB lower than that obtained by our CPM. The reason is that our CPM (strip operation with prediction network) can capture more relevant information from adjacent blocks for prediction with the strip operation.

\subsubsection{Comparison to Postprocessing Networks}
To verify the effectiveness of our BPM, we compared our method with state-of-the-art (SOTA) postprocessing methods, including DHDN~\cite{park2019densely}, GRDN~\cite{kim2019grdn} and CLIC19\_MS ~\cite{lu2019learned}. All of the above methods are retrained on the same dataset and tested on the Kodak dataset. As shown in Table~\ref{tab:post}, our BPM is superior to existing works, achieving approximately 0.1 dB gain compared with the GRDN~\cite{kim2019grdn}. To demonstrate the visual difference of decoded images before and after the BPM, we visualize the decoded images, which are encoded with the model optimized for MSE with $\lambda=192$. As shown in Fig.~\ref{fig:vsbpm}, due to the effect of the padding operation, reconstructed edge pixels of neighbor blocks contain chromatism, which results in blocking effects that greatly affect the subjective quality. Through utilizing the information between blocks, our BPM module can smooth out the sudden changes in edges and recover some information that has been lost during the transformation (e.g., eyepit), which achieves better subjective quality.

\subsubsection{Ablation on proposed modules and block size}
To analyze the contributions of CPM and BPM in our work, we utilize the LBHIC model without the CPM and BPM as our baseline model, which means that we compress and reconstruct the blocks directly after block partition. We then investigate the effectiveness of the CPM and BPM by adding them to the baseline model separately. As shown in Fig.~\ref{fig:ab1}, the CPM (baseline with CPM) and BPM (baseline with BPM) can produce comparable compression performance improvement compared with the baseline. With the combination of CPM and BPM (our method), we can further improve the compression performance compared to that with only CPM or BPM.

We also investigate the effects of block sizes on the compression performance in Fig.~\ref{fig:ab}(b). Specifically, we utilize VTM as the anchor, and calculate the BD rate between the VTM and different methods in Fig.~\ref{fig:ab}(b). The impacts of block size on learned block-based coding can be summarized into three aspects: 1) performance loss caused by block partition, 2) performance gain caused by block prediction, and 3) performance gain obtained by removing block artifacts. For block partition, the smaller block will bring worse performance for the existing learned codecs, because the correlation between blocks is destroyed. As for the block prediction, its performance will improve as the block becomes smaller, because the prediction is easier on smaller blocks. Since smaller blocks tend to have more blockiness, the gain obtained by post-processing on smaller blocks tend to be more obvious. Due to the influence of above three aspects, the factors affecting the final performance actually depend on the impact degree of these three aspects. Hence, in Fig.~\ref{fig:ab}(b), we conduct further experiments to reveal the influence of the above three aspects. 
It can be seen that due to the influence of block partition, the baseline models (i.e, w/o CPM and BPM) of different blocks size have different performances and smaller blocks cause worse performance due to the impact of block partition. Meanwhile, the relative gain of prediction and post-processing (i.e, Baseline+CPM and Baseline+CPM+BPM) increases as the block becomes smaller. However, due to the greater impact of block partition on learned image compression, in the end, smaller blocks cannot achieve better performance.  Furthermore, we can notice that our LBHIC (i.e, Baseline+CPM+BPM) with block size 128 and 256 can achieve similar BD-rate.  Nevertheless, as the block size increases, the performance gain will steadily decrease, and the degree of parallelism also decreases substantially. Considering a trade-off of coding performance and parallelism, we finally use 128 as our block size in our LBHIC.

\subsubsection{Proportion of different modules in coding time.}
\begin{table}[t]
\caption{Proportion of different modules in coding time.}
\centering
\setlength{\tabcolsep}{0.01\linewidth}{
\begin{tabular}{ccccccc}
\toprule
    & CPM   & Transformation & \begin{tabular}[c]{@{}c@{}}Encode\\ Entropy\end{tabular} & \begin{tabular}[c]{@{}c@{}}Inverse\\ Transformation\end{tabular} & \begin{tabular}[c]{@{}c@{}}Decode\\ Entropy\end{tabular} & BPM   \\ \midrule
    \begin{tabular}[c]{@{}c@{}}Percen-\\ -tage \end{tabular} & 3.79\%  & 1.40\% & 8.19\%     & 1.46\%   & 69.02\%     & 16.14\% \\ \bottomrule
\end{tabular}}
\label{tab:ratiotime}
\end{table}
We analyze the proportion of different modules in the compression network to the total encoding and decoding time, and the results are given in Table~\ref{tab:ratiotime}.  Since the context entropy model at the decoder requires element-by-element decoding in a auto-regressive manner, it occupies most of the coding time, which promotes the birth of our block-based coding scheme. Because the block-based scheme can alleviate this problem through a parallel strategy.

\section{Conclusion}
\label{sec:5}
In this paper, we reconcile compression efficiency with rate-distortion performance when fostering progress in developing a learning-based image compression framework. We inherit the advantages of the traditional image codecs in the proposed learned block-based hybrid image compression scheme (LBHIC) to improve both the efficiency and effectiveness. Our proposed framework is composed of a block partition, contextual prediction module (CPM), and boundary-aware postprocessing module (BPM). Among these components, the block partition activates the possibility of acceleration, CPM effectively utilizes the correlation between blocks to improve coding efficiency, and BPM takes into account the block effect to improve the subjective and objective quality. The experimental results show that our approach not only achieves SOTA performance but also provides almost 10x improvement in decoded runtime, yielding a high-performance, efficiently learned image codec.

\ifCLASSOPTIONcaptionsoff
  \newpage
\fi

\bibliographystyle{./bibtex/IEEEtran}
\bibliography{./bibtex/HLF.bib}

\begin{thebibliography}{10}
\providecommand{\url}[1]{#1}
\csname url@samestyle\endcsname
\providecommand{\newblock}{\relax}
\providecommand{\bibinfo}[2]{#2}
\providecommand{\BIBentrySTDinterwordspacing}{\spaceskip=0pt\relax}
\providecommand{\BIBentryALTinterwordstretchfactor}{4}
\providecommand{\BIBentryALTinterwordspacing}{\spaceskip=\fontdimen2\font plus
\BIBentryALTinterwordstretchfactor\fontdimen3\font minus
  \fontdimen4\font\relax}
\providecommand{\BIBforeignlanguage}[2]{{%
\expandafter\ifx\csname l@#1\endcsname\relax
\typeout{** WARNING: IEEEtran.bst: No hyphenation pattern has been}%
\typeout{** loaded for the language `#1'. Using the pattern for}%
\typeout{** the default language instead.}%
\else
\language=\csname l@#1\endcsname
\fi
#2}}
\providecommand{\BIBdecl}{\relax}
\BIBdecl

\bibitem{wallace1992jpeg}
G.~K. Wallace, ``The jpeg still picture compression standard,'' \emph{IEEE
  transactions on consumer electronics}, vol.~38, no.~1, pp. xviii--xxxiv,
  1992.

\bibitem{bellard2015bpg}
F.~Bellard, ``Bpg image format,'' \emph{URL https://bellard. org/bpg}, vol.~1,
  2015.

\bibitem{bross2018versatile}
B.~Bross, J.~Chen, and S.~Liu, ``Versatile video coding (draft 5),''
  \emph{JVET-K1001}, 2018.

\bibitem{schiopu2019cnn}
I.~Schiopu, H.~Huang, and A.~Munteanu, ``Cnn-based intra-prediction for
  lossless hevc,'' \emph{IEEE Transactions on Circuits and Systems for Video
  Technology}, vol.~30, no.~7, pp. 1816--1828, 2019.

\bibitem{balle2017end}
J.~Ball{\'e}, V.~Laparra, and E.~P. Simoncelli, ``End-to-end optimized image
  compression,'' in \emph{International Conference on Learning Representations
  (ICLR)}, 2017.

\bibitem{balle2018variational}
J.~Ball{\'e}, D.~Minnen, S.~Singh, S.~J. Hwang, and N.~Johnston, ``Variational
  image compression with a scale hyperprior,'' in \emph{International
  Conference on Learning Representations (ICLR)}, 2018.

\bibitem{minnen2018joint}
D.~Minnen, J.~Ball{\'e}, and G.~D. Toderici, ``Joint autoregressive and
  hierarchical priors for learned image compression,'' in \emph{Advances in
  Neural Information Processing Systems}, 2018, pp. 10\,794--10\,803.

\bibitem{johnston2018improved}
N.~Johnston, D.~Vincent, D.~Minnen, M.~Covell, S.~Singh, T.~Chinen,
  S.~Jin~Hwang, J.~Shor, and G.~Toderici, ``Improved lossy image compression
  with priming and spatially adaptive bit rates for recurrent networks,'' in
  \emph{Proceedings of the IEEE Conference on Computer Vision and Pattern
  Recognition}, 2018, pp. 4385--4393.

\bibitem{johnston2019computationally}
N.~Johnston, E.~Eban, A.~Gordon, and J.~Ball{\'e}, ``Computationally efficient
  neural image compression,'' \emph{arXiv preprint arXiv:1912.08771}, 2019.

\bibitem{sun2020semantic}
S.~Sun, T.~He, and Z.~Chen, ``Semantic structured image coding framework for
  multiple intelligent applications,'' \emph{IEEE Transactions on Circuits and
  Systems for Video Technology}, 2020.

\bibitem{chen2019learning}
Z.~Chen and T.~He, ``Learning based facial image compression with semantic
  fidelity metric,'' \emph{Neurocomputing}, vol. 338, pp. 16--25, 2019.

\bibitem{zhang2019learned}
Z.~Zhang, Z.~Chen, J.~Lin, and W.~Li, ``Learned scalable image compression with
  bidirectional context disentanglement network,'' in \emph{2019 IEEE
  International Conference on Multimedia and Expo (ICME)}.\hskip 1em plus 0.5em
  minus 0.4em\relax IEEE, 2019, pp. 1438--1443.

\bibitem{guo2021causal}
Z.~Guo, Z.~Zhang, R.~Feng, and Z.~Chen, ``Causal contextual prediction for
  learned image compression,'' \emph{IEEE Transactions on Circuits and Systems
  for Video Technology}, 2021.

\bibitem{mentzer2020high}
F.~Mentzer, G.~D. Toderici, M.~Tschannen, and E.~Agustsson, ``High-fidelity
  generative image compression,'' \emph{Advances in Neural Information
  Processing Systems}, vol.~33, 2020.

\bibitem{franzen1999kodak}
R.~Franzen, ``Kodak lossless true color image suite,'' \emph{source:
  http://r0k. us/graphics/kodak}, vol.~4, no.~2, 1999.

\bibitem{hou2020strip}
Q.~Hou, L.~Zhang, M.-M. Cheng, and J.~Feng, ``Strip pooling: Rethinking spatial
  pooling for scene parsing,'' in \emph{Proceedings of the IEEE/CVF Conference
  on Computer Vision and Pattern Recognition}, 2020, pp. 4003--4012.

\bibitem{toderici2015variable}
G.~Toderici, S.~M. O'Malley, S.~J. Hwang, D.~Vincent, D.~Minnen, S.~Baluja,
  M.~Covell, and R.~Sukthankar, ``Variable rate image compression with
  recurrent neural networks,'' in \emph{International Conference on Learning
  Representations (ICLR)}, 2016.

\bibitem{toderici2017full}
G.~Toderici, D.~Vincent, N.~Johnston, S.~Jin~Hwang, D.~Minnen, J.~Shor, and
  M.~Covell, ``Full resolution image compression with recurrent neural
  networks,'' in \emph{Proceedings of the IEEE Conference on Computer Vision
  and Pattern Recognition}, 2017, pp. 5306--5314.

\bibitem{balle2016end}
J.~Ball{\'e}, V.~Laparra, and E.~P. Simoncelli, ``End-to-end optimization of
  nonlinear transform codes for perceptual quality,'' in \emph{Picture Coding
  Symposium (PCS), 2016}.\hskip 1em plus 0.5em minus 0.4em\relax IEEE, 2016,
  pp. 1--5.

\bibitem{chen2019neural}
T.~Chen, H.~Liu, Z.~Ma, Q.~Shen, X.~Cao, and Y.~Wang, ``Neural image
  compression via non-local attention optimization and improved context
  modeling,'' \emph{arXiv preprint arXiv:1910.06244}, 2019.

\bibitem{qian2020learning}
Y.~Qian, Z.~Tan, X.~Sun, M.~Lin, D.~Li, Z.~Sun, H.~Li, and R.~Jin, ``Learning
  accurate entropy model with global reference for image compression,''
  \emph{arXiv preprint arXiv:2010.08321}, 2020.

\bibitem{agustsson2017soft}
E.~Agustsson, F.~Mentzer, M.~Tschannen, L.~Cavigelli, R.~Timofte, L.~Benini,
  and L.~V. Gool, ``Soft-to-hard vector quantization for end-to-end learning
  compressible representations,'' in \emph{Advances in Neural Information
  Processing Systems}, 2017, pp. 1141--1151.

\bibitem{agustsson2020universally}
E.~Agustsson and L.~Theis, ``Universally quantized neural compression,''
  \emph{arXiv preprint arXiv:2006.09952}, 2020.

\bibitem{mentzer2018conditional}
F.~Mentzer, E.~Agustsson, M.~Tschannen, R.~Timofte, and L.~Van~Gool,
  ``Conditional probability models for deep image compression,'' in
  \emph{Proceedings of the IEEE Conference on Computer Vision and Pattern
  Recognition}, 2018, pp. 4394--4402.

\bibitem{Lee2019Context}
J.~Lee, S.~Cho, and S.-K. Beack, ``Context-adaptive entropy model for
  end-to-end optimized image compression,'' in \emph{the 7th Int. Conf. on
  Learning Representations}, May 2019.

\bibitem{guo20203}
Z.~Guo, Y.~Wu, R.~Feng, Z.~Zhang, and Z.~Chen, ``3-d context entropy model for
  improved practical image compression,'' in \emph{Proceedings of the IEEE/CVF
  Conference on Computer Vision and Pattern Recognition Workshops}, 2020, pp.
  116--117.

\bibitem{hu2020coarse}
Y.~Hu, W.~Yang, and J.~Liu, ``Coarse-to-fine hyper-prior modeling for learned
  image compression.'' in \emph{AAAI}, 2020, pp. 11\,013--11\,020.

\bibitem{dabov2007image}
K.~Dabov, A.~Foi, V.~Katkovnik, and K.~Egiazarian, ``Image denoising by sparse
  3-d transform-domain collaborative filtering,'' \emph{IEEE Transactions on
  image processing}, vol.~16, no.~8, pp. 2080--2095, 2007.

\bibitem{zhang2017beyond}
K.~Zhang, W.~Zuo, Y.~Chen, D.~Meng, and L.~Zhang, ``Beyond a gaussian denoiser:
  Residual learning of deep cnn for image denoising,'' \emph{IEEE Transactions
  on Image Processing}, vol.~26, no.~7, pp. 3142--3155, 2017.

\bibitem{kupyn2018deblurgan}
O.~Kupyn, V.~Budzan, M.~Mykhailych, D.~Mishkin, and J.~Matas, ``Deblurgan:
  Blind motion deblurring using conditional adversarial networks,'' in
  \emph{Proceedings of the IEEE Conference on Computer Vision and Pattern
  Recognition}, 2018, pp. 8183--8192.

\bibitem{kupyn2019deblurgan}
O.~Kupyn, T.~Martyniuk, J.~Wu, and Z.~Wang, ``Deblurgan-v2: Deblurring
  (orders-of-magnitude) faster and better,'' in \emph{Proceedings of the IEEE
  International Conference on Computer Vision}, 2019, pp. 8878--8887.

\bibitem{li2020learning}
X.~Li, X.~Jin, J.~Lin, S.~Liu, Y.~Wu, T.~Yu, W.~Zhou, and Z.~Chen, ``Learning
  disentangled feature representation for hybrid-distorted image restoration,''
  in \emph{European Conference on Computer Vision}.\hskip 1em plus 0.5em minus
  0.4em\relax Springer, 2020, pp. 313--329.

\bibitem{lee2019hybrid}
J.~Lee, S.~Cho, and M.~Kim, ``A hybrid architecture of jointly learning image
  compression and quality enhancement with improved entropy minimization,''
  \emph{arXiv preprint arXiv:1912.12817}, 2019.

\bibitem{jin2020dual}
Z.~Jin, M.~Z. Iqbal, W.~Zou, X.~Li, and E.~Steinbach, ``Dual-stream multi-path
  recursive residual network for jpeg image compression artifacts reduction,''
  \emph{IEEE Transactions on Circuits and Systems for Video Technology},
  vol.~31, no.~2, pp. 467--479, 2020.

\bibitem{kim2020towards}
Y.~Kim, S.~Cho, J.~Lee, S.-Y. Jeong, J.~Soo~Choi, and J.~Do, ``Towards the
  perceptual quality enhancement of low bit-rate compressed images,'' in
  \emph{Proceedings of the IEEE/CVF Conference on Computer Vision and Pattern
  Recognition Workshops}, 2020, pp. 136--137.

\bibitem{li2020multi}
X.~Li, S.~Sun, Z.~Zhang, and Z.~Chen, ``Multi-scale grouped dense network for
  vvc intra coding,'' in \emph{Proceedings of the IEEE/CVF Conference on
  Computer Vision and Pattern Recognition Workshops}, 2020, pp. 158--159.

\bibitem{wang2018esrgan}
X.~Wang, K.~Yu, S.~Wu, J.~Gu, Y.~Liu, C.~Dong, Y.~Qiao, and C.~Change~Loy,
  ``Esrgan: Enhanced super-resolution generative adversarial networks,'' in
  \emph{Proceedings of the European Conference on Computer Vision (ECCV)},
  2018, pp. 0--0.

\bibitem{chen2019learningvideo}
Z.~Chen, T.~He, X.~Jin, and F.~Wu, ``Learning for video compression,''
  \emph{IEEE Transactions on Circuits and Systems for Video Technology},
  vol.~30, no.~2, pp. 566--576, 2019.

\bibitem{lin2020spatial}
C.~Lin, J.~Yao, F.~Chen, and L.~Wang, ``A spatial rnn codec for end-to-end
  image compression,'' in \emph{Proceedings of the IEEE/CVF Conference on
  Computer Vision and Pattern Recognition}, 2020, pp. 13\,269--13\,277.

\bibitem{sullivan2012overview}
G.~J. Sullivan, J.-R. Ohm, W.-J. Han, and T.~Wiegand, ``Overview of the high
  efficiency video coding (hevc) standard,'' \emph{IEEE Transactions on
  circuits and systems for video technology}, vol.~22, no.~12, pp. 1649--1668,
  2012.

\bibitem{wang2018non}
X.~Wang, R.~Girshick, A.~Gupta, and K.~He, ``Non-local neural networks,'' in
  \emph{Proceedings of the IEEE conference on computer vision and pattern
  recognition}, 2018, pp. 7794--7803.

\bibitem{ronneberger2015u}
O.~Ronneberger, P.~Fischer, and T.~Brox, ``U-net: Convolutional networks for
  biomedical image segmentation,'' in \emph{International Conference on Medical
  image computing and computer-assisted intervention}.\hskip 1em plus 0.5em
  minus 0.4em\relax Springer, 2015, pp. 234--241.

\bibitem{lu2019learned}
M.~Lu, T.~Chen, H.~Liu, and Z.~Ma, ``Learned image restoration for vvc intra
  coding.'' in \emph{CVPR Workshops}, 2019, p.~0.

\bibitem{zhang2018residual}
Y.~Zhang, Y.~Tian, Y.~Kong, B.~Zhong, and Y.~Fu, ``Residual dense network for
  image super-resolution,'' in \emph{Proceedings of the IEEE conference on
  computer vision and pattern recognition}, 2018, pp. 2472--2481.

\bibitem{kingma2014adam}
D.~P. Kingma and J.~Ba, ``Adam: A method for stochastic optimization,''
  \emph{arXiv preprint arXiv:1412.6980}, 2014.

\bibitem{kim2019grdn}
D.-W. Kim, J.~Ryun~Chung, and S.-W. Jung, ``Grdn: Grouped residual dense
  network for real image denoising and gan-based real-world noise modeling,''
  in \emph{Proceedings of the IEEE Conference on Computer Vision and Pattern
  Recognition Workshops}, 2019, pp. 0--0.

\bibitem{deng2009imagenet}
J.~Deng, W.~Dong, R.~Socher, L.-J. Li, K.~Li, and L.~Fei-Fei, ``Imagenet: A
  large-scale hierarchical image database,'' in \emph{2009 IEEE conference on
  computer vision and pattern recognition}.\hskip 1em plus 0.5em minus
  0.4em\relax Ieee, 2009, pp. 248--255.

\bibitem{agustsson2017ntire}
E.~Agustsson and R.~Timofte, ``Ntire 2017 challenge on single image
  super-resolution: Dataset and study,'' in \emph{Proceedings of the IEEE
  Conference on Computer Vision and Pattern Recognition Workshops}, 2017, pp.
  126--135.

\bibitem{asuni2014testimages}
N.~Asuni and A.~Giachetti, ``Testimages: a large-scale archive for testing
  visual devices and basic image processing algorithms.'' in \emph{STAG}, 2014,
  pp. 63--70.

\bibitem{wang2003multiscale}
Z.~Wang, E.~P. Simoncelli, and A.~C. Bovik, ``Multiscale structural similarity
  for image quality assessment,'' in \emph{The Thrity-Seventh Asilomar
  Conference on Signals, Systems \& Computers, 2003}, vol.~2.\hskip 1em plus
  0.5em minus 0.4em\relax Ieee, 2003, pp. 1398--1402.

\bibitem{cheng2020learned}
Z.~Cheng, H.~Sun, M.~Takeuchi, and J.~Katto, ``Learned image compression with
  discretized gaussian mixture likelihoods and attention modules,'' in
  \emph{Proceedings of the IEEE/CVF Conference on Computer Vision and Pattern
  Recognition}, 2020, pp. 7939--7948.

\bibitem{bjontegaard2001calculation}
G.~Bjontegaard, ``Calculation of average psnr differences between rd-curves,''
  \emph{VCEG-M33}, 2001.

\bibitem{dumas2019context}
T.~Dumas, A.~Roumy, and C.~Guillemot, ``Context-adaptive neural network-based
  prediction for image compression,'' \emph{IEEE Transactions on Image
  Processing}, vol.~29, pp. 679--693, 2019.

\bibitem{hu2019progressive}
Y.~Hu, W.~Yang, M.~Li, and J.~Liu, ``Progressive spatial recurrent neural
  network for intra prediction,'' \emph{IEEE Transactions on Multimedia},
  vol.~21, no.~12, pp. 3024--3037, 2019.

\bibitem{richardson2004h}
I.~E. Richardson, \emph{H. 264 and MPEG-4 video compression: video coding for
  next-generation multimedia}.\hskip 1em plus 0.5em minus 0.4em\relax John
  Wiley \& Sons, 2004.

\bibitem{park2019densely}
B.~Park, S.~Yu, and J.~Jeong, ``Densely connected hierarchical network for
  image denoising,'' in \emph{Proceedings of the IEEE Conference on Computer
  Vision and Pattern Recognition Workshops}, 2019, pp. 0--0.

\end{thebibliography}





\end{document}